%% file: main.tex
\documentclass[sigconf,natbib,anonymous=false,review=false,screen=true]{acmart}

\acmSubmissionID{352}

\input{packages}
\input{definitions}
\input{authors}
\input{meta}

\begin{document}
\fancyhead{}
\fancyfoot{}

\title{Cross-Domain Contract Element Extraction with a \\ Bi-directional Feedback Clause-Element Relation Network}

\input{sections/abstract.tex}



\maketitle

\acresetall

\input{sections/introduction.tex}
\input{sections/related_work.tex}
\input{sections/method.tex}
\input{sections/experiments.tex}
\input{sections/results.tex}
\input{sections/analysis.tex}
\input{sections/conclusion.tex}

\section*{Reproducibility}
To facilitate reproducibility of the results reported in this paper, 
the code and data used are available at \url{https://github.com/WZH-NLP/Bi-FLEET}.

\input{sections/acknowledgement.tex}

\clearpage
\bibliographystyle{ACM-Reference-Format}
\balance
\bibliography{main}

\end{document}

%% file: packages.tex
\usepackage{booktabs} 
\usepackage{algorithm, algpseudocode}
\usepackage{amsmath}
\usepackage{graphics}
\usepackage{epsfig}
\usepackage{graphicx}
\usepackage{xcolor}
\usepackage[skip=0pt]{caption}
\usepackage{subfigure}
\usepackage{balance}
\usepackage{multirow}
\usepackage{mathrsfs}
\usepackage{acronym}
\usepackage{placeins}
\usepackage{xcolor}
\usepackage[inline]{enumitem}
\usepackage{fancyhdr}

%% file: definitions.tex
\AtBeginDocument{%
  \providecommand\BibTeX{{%
    \normalfont B\kern-0.5em{\scshape i\kern-0.25em b}\kern-0.8em\TeX}}}

\acrodef{CEE}{contract element extraction}
\acrodef{CC}{clause classification}
\acrodef{NER}{named entity recognition}
\acrodef{Bi-FLEET}{Bi-directional Feedback cLause-Element relaTion network}
\acrodef{C-E}{clause-element}

\newcommand{\smallnegskip}{\vspace*{-2mm}}
\newcommand{\negskip}{\vspace*{-3mm}}

%% file: authors.tex
\author{Zihan Wang$^1$\qquad Hongye Song$^2$\qquad Zhaochun Ren$^1$*\qquad Pengjie Ren$^1$\qquad Zhumin Chen$^1$}
\author{
Xiaozhong Liu$^3$\qquad Hongsong Li$^2$ \qquad Maarten de Rijke$^{4,5}$
}

\makeatletter
\def\authornotetext#1{
\if@ACM@anonymous\else
    \g@addto@macro\@authornotes{
    \stepcounter{footnote}\footnotetext{#1}}
\fi}
\makeatother
\authornotetext{Corresponding author.}

\affiliation{
 \institution{\textsuperscript{\rm 1}Shandong University, Qingdao \country{China}}
  \institution{\textsuperscript{\rm 2}Alibaba Group, Hangzhou, China}
 \institution{\textsuperscript{\rm 3}Indiana University Bloomington, Bloomington \country{United States}}
 \institution{\textsuperscript{\rm 4}University of Amsterdam, Amsterdam \country{The Netherlands}\quad \textsuperscript{\rm 5}Ahold Delhaize Research, Zaandam \country{The Netherlands}}
}

\email{zihanwang.sdu@gmail.com, {hongye.shy, hongsong.lhs}@alibaba-inc.com, {zhaochun.ren, chenzhumin}@sdu.edu.cn} 
\email{jay.ren@outlook.com, liu237@indiana.edu, m.derijke@uva.nl}

\def\authors{Zihan Wang, Hongye Song, Zhaochun Ren, Pengjie Ren, Zhumin Chen, Xiaozhong Liu, Hongsong Li, and Maarten de Rijke}

\renewcommand{\shortauthors}{Trovato and Tobin, et al.}

%% file: meta.tex
\copyrightyear{2021}
\acmYear{2021}
\setcopyright{acmlicensed}\acmConference[SIGIR '21]{Proceedings of the 44th International ACM SIGIR Conference on Research and Development in Information Retrieval}{July 11--15, 2021}{Virtual Event, Canada}
\acmBooktitle{Proceedings of the 44th International ACM SIGIR Conference on Research and Development in Information Retrieval (SIGIR '21), July 11--15, 2021, Virtual Event, Canada}
\acmPrice{15.00}
\acmDOI{10.1145/3404835.3462873}
\acmISBN{978-1-4503-8037-9/21/07}

%% file: sections/abstract.tex

\begin{abstract}
Contract element extraction (CEE) is the novel task of automatically identifying and extracting legally relevant elements such as contract dates, payments, and legislation references from contracts.
Automatic methods for this task view it as a sequence labeling problem and dramatically reduce human labor. However, as contract genres and element types may vary widely, a significant challenge for this sequence labeling task is how to transfer knowledge from one domain to another, i.e., cross-domain CEE.
Cross-domain CEE differs from cross-domain \acf{NER} in two important ways. First, contract elements are far more fine-grained than named entities, which hinders the transfer of extractors. Second, the extraction zones for cross-domain \ac{CEE} are much larger than for cross-domain \ac{NER}. As a result, the contexts of elements from different domains can be more diverse.

We propose a framework, the \acfi{Bi-FLEET}, for the cross-domain \ac{CEE} task that addresses the above challenges. \ac{Bi-FLEET} has three main components: 
\begin{enumerate*}
\item a context encoder, 
\item a clause-element relation encoder, and 
\item an inference layer. 
\end{enumerate*}
To incorporate invariant knowledge about element and clause types, a clause-element graph is constructed across domains and a hierarchical graph neural network is adopted in the clause-element relation encoder. 
To reduce the influence of context variations, a multi-task framework with a bi-directional feedback scheme is designed in the inference layer, conducting both clause classification and element extraction. 
The experimental results over both cross-domain \ac{NER} and \ac{CEE} tasks show that \ac{Bi-FLEET} significantly outperforms state-of-the-art baselines. 
\end{abstract}

%% file: sections/introduction.tex

\section{Introduction}
\begin{figure}
  \centering
  \includegraphics[width=\linewidth]{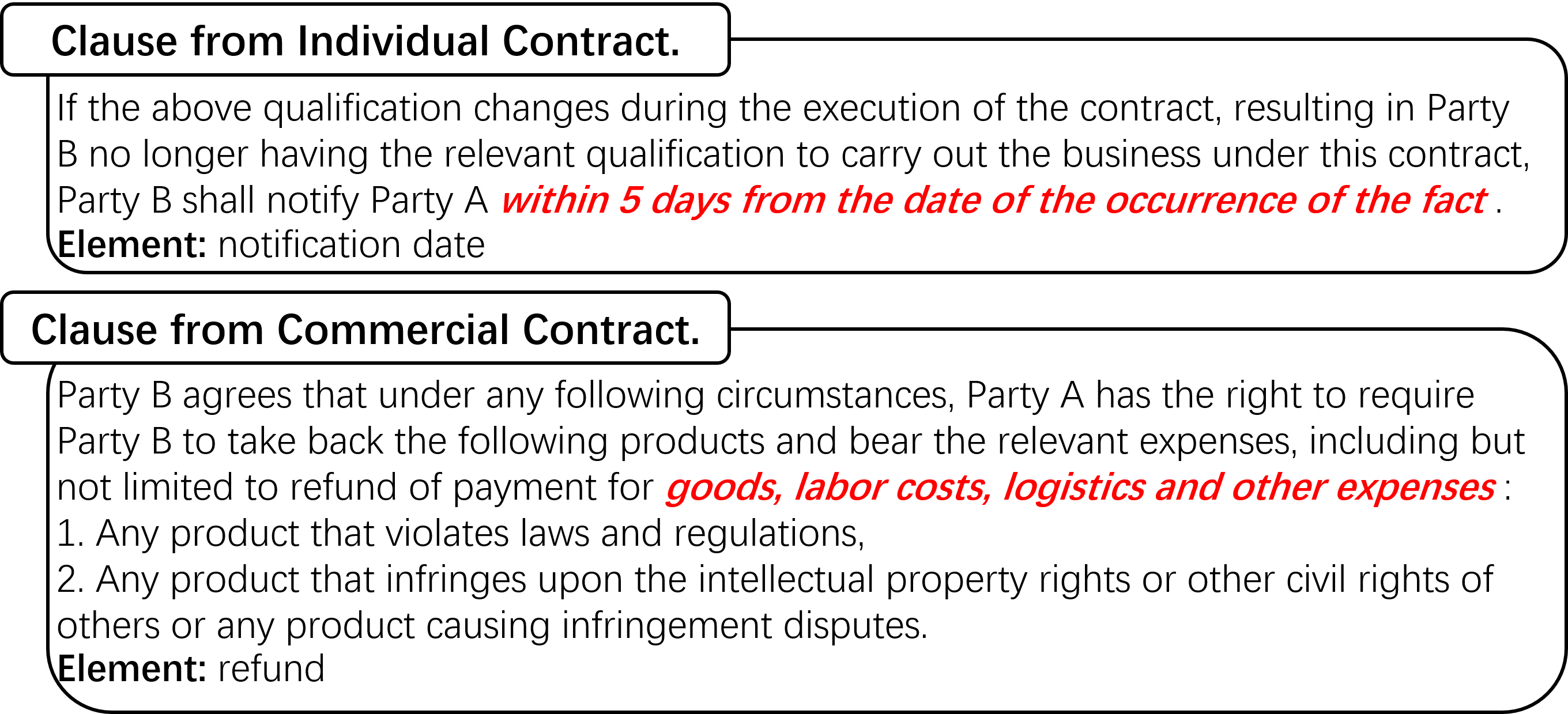}
  \caption{Examples of contract element extractions. (Top): A clause from an individual contract. (Bottom): A clause from a commercial contract.}
  \label{fig:CEE examples} 
\end{figure}

Extracting information from contracts or other legal agreements is a fundamental task for businesses around the world~\citep{chalkidis2019neural,chalkidis2017deep,milosevic2004design}. 
Many thousands of contracts, relevant to a large variety of transactions (such as loans, investments, or leases), are drawn up every day. These contracts usually contain legally relevant elements such as termination dates or contract parties. 
Manually monitoring the legally relevant information in a large number of contracts is time-consuming, labor-intensive, and error-prone, putting a heavy burden on law firms, companies, and government agencies~\citep{sun2019toi}. 
Automatic \acfi{CEE} is increasingly attracting interest~\citep{DBLP:conf/jurix/MaxwellS08}). 
As Fig.~\ref{fig:CEE examples}(top) illustrates, given a clause from a contract, the goal of the \ac{CEE} task is to find the legally relevant elements (such as ``within 5 days \ldots\ fact'') in the contract. \ac{CEE} can
bring useful insights into contracts and it can facilitate downstream applications, such as relevant clause retrieval or risk assessment~\citep{DBLP:conf/emnlp/BorchmannWGKJSP20}.

Modern information extraction methods consider the \ac{CEE} task as a sequence labeling problem, classifying each word as a (part of a) type of contract element~\citep{chalkidis2019neural, chalkidis2017deep, sun2019toi}. 
The \ac{CEE} task is challenging due to large variations in element mentions. 
Another important challenge for this sequence labeling problem is how to transfer knowledge from one domain to another~\citep{DBLP:conf/iclr/YangSC17}.
For example, as Fig.~\ref{fig:CEE examples} shows, compared with individual contracts (top), commercial contracts (bottom) are more formal, giving more precise and complex explanations in the contract clauses.
Because of the differences in context genres and element types, transferring an element extractor from the individual domain to the commercial domain (or vice versa) is a challenging problem. 

In this paper, we focus on cross-domain \acf{CEE}.
\begin{table}[t]
  \small
  \centering
  \caption{Examples of the most frequent elements in clauses from different contract domains. Percentages in brackets indicate ratios of elements appearing in clauses from the individual/commercial contracts.}
  \label{tab:examples of C-E relations}
  \begin{tabular}{ll}
  \toprule
    Clause & The most frequent elements (individual/commercial) \\
  \midrule
    Payment & Payment period (23.8\%/25.6\%) and rate (12.3\%/10.9\%) \\
  \midrule
    Deposit & Deposit rate (26.4\%/40.6\%) and amount (12.4\%/14.5\%)\\
  \midrule
    Effectiveness & Effective mode (45.5\%/31.3\%) and time (17.7\%/26.7\%) \\
  \bottomrule
\end{tabular}
\end{table}
Cross-domain \ac{CEE} is similar to the cross-domain \ac{NER} task in many ways~\citep{chalkidis2017deep, chalkidis2017extracting}. Both tasks aim to identify key information (elements or entities) from texts in different domains and have similar problem formulations (discussed in detail in Sec.~\ref{subsec:problem formulation}). 
However, existing cross-domain \ac{NER} methods are not directly applicable to the cross-domain \ac{CEE} task and fail to address two important challenges:

\begin{description}[leftmargin=3.5mm,nosep]
\item[Challenge 1: Transfer of fine-grained element types.] 
\sloppy A ge\-ne\-ric named entity recognizer typically extracts only several types of entities, such as persons, organizations, locations, dates or amounts. 
However, in the \ac{CEE} task, there are far more types of fine-grained contract elements (e.g., more than 70 in our dataset). 
For example, a typical entity recognizer may identify dates, but cannot distinguish between start, termination or other dates. 
In the cross-domain setting, the presence of a larger number of fine-grained contract elements makes it more difficult to transfer the extractor. 
To facilitate the transfer process, invariant knowledge about element types should be incorporated. 
As Table~\ref{tab:examples of C-E relations} shows, dependencies between clause and element types are usually shared across domains. 
E.g., the payment period  usually exists in the payment clauses, but not the deposit clauses, which is independent of contract domains.

\item[Challenge 2: Transfer of larger extraction zones.]
\sloppy The \ac{NER} task mainly focuses on locating entities in a single sentence. In contrast, an extractor for the \ac{CEE} task needs to find contract elements from multiple sentences in a clause. 
That is, extraction zones (clauses) for contract elements are much larger than the contexts of entities.
Sentences in clauses can be organized in various ways for different domains, which brings a new challenge for the cross-domain \ac{CEE} task. To reduce the influence of sentence organizations, clause types should be decided first.
Besides, classifying clauses is of great importance for extracting contract elements. 
E.g., if we know a clause is a loan clause, it is easier to figure out that the dates in the clause are  loan or repayment dates. 
Hence, conducting clause classification and element extraction at the same time can enhance the inference process.
\end{description}

\noindent%
To address the challenges identified above, we propose a framework, named the \acfi{Bi-FLEET}, that captures shared knowledge about element types in two domains and interactions between contract element extraction and clause classification. 
As Fig.~\ref{fig:framework} shows, \ac{Bi-FLEET} is composed of three main components: 
\begin{enumerate*}[label=(\arabic*)]
\item a context encoder, 
\item a clause-element relation encoder, and 
\item an inference layer. 
\end{enumerate*}
First, the context encoder embeds all sentences in clauses. Then, to capture invariant relations between element and clause types, in the clause-element relation encoder, a \acf{C-E} graph is constructed across domains and a hierarchical graph neural network (GNN) is implemented to encode the shared clause-element dependencies. Finally, to identify the clause types and facilitate the element extraction process, a multi-task framework with a bi-directional feedback scheme is proposed in the inference layer. 
Similar to the human annotation process, both clause classification and contract element extraction tasks are conducted. To model interactions between tasks and improve the overall performance, forward and backward information is calculated using representations of contexts and types from the context and clause-element relation encoders, respectively. 
To evaluate our proposed framework to extract contract elements in cross-domain CEE, we collect both individual and commercial contracts to establish a cross-domain CEE dataset.
Experimental results using both cross-domain \ac{NER} and \ac{CEE} datasets demonstrate the effectiveness of \ac{Bi-FLEET}.

The contributions of this paper can be summarized as follows:
\begin{itemize}[leftmargin=*,nosep]
    \item To the best of our knowledge, ours is the first work on the cross-domain \ac{CEE} task. Compared with cross-domain \ac{NER}, more fine-grained element types and larger extraction zones bring critical challenges for the transfer process.
    \item We propose a framework, named the \acfi{Bi-FLEET}, to capture invariant clause-element dependencies and interactions between contract element extraction and clause classification.
    \item We establish a cross-domain \ac{CEE} dataset by collecting individual contracts and commercial contracts. To the best of our knowledge, this is the first dataset for cross-domain \ac{CEE}.
    \item Experimental results show that the proposed \ac{Bi-FLEET} model achieves significant improvements over baselines in both cross-domain \ac{NER} and \ac{CEE} tasks. 
\end{itemize}

%% file: sections/related_work.tex

\section{Related work}
\label{sec:related work}
We survey related work along three dimensions: 
\begin{enumerate*}
\item legal information retrieval and extraction,
\item contract element extraction, and
\item cross-domain named entity recognition.
\end{enumerate*}

\subsection{Legal information retrieval and extraction}
The digitization of legal documents has given rise to the development of legal information retrieval systems~\citep{DBLP:conf/jurix/MaxwellS08, DBLP:conf/coling/KienNBTNP20}. Numerous challenges for legal information retrieval and extraction have been presented~\citep{DBLP:journals/jd/Chu11,DBLP:conf/icail/KanoKGS17}. Traditional legal search systems are keyword-based, relying heavily on the professional knowledge of end-users. Recently, to alleviate the dependence on user's expertise and improve retrieval effectiveness, a significant amount of effort~\citep{do2017legal, perera2018legal, DBLP:journals/ail/TranNTS20, DBLP:conf/coling/KienNBTNP20} has been devoted to the automatic classification of legal documents and queries. The extraction of key legal concepts can also facilitate the retrieval process~\citep{DBLP:conf/icail/TranNS19, DBLP:conf/emnlp/BorchmannWGKJSP20}.

To the best of our knowledge, there is no previous work on legal information retrieval and extraction that focuses on cross-domain \acf{CEE}.  

\subsection{Contract element extraction}
The goal of \ac{CEE} is to recognize essential legal elements, such as execution date, jurisdiction, and amount, in legal documents~\citep{curtotti2010corpus,indukuri2010mining}.
Early \ac{CEE} methods are mainly rule-based or traditional statistical methods.  
\citet{chalkidis2017extracting} introduce 11 contract element types and extract contract elements based on Logistic Regression and SVM with hand-crafted features. 
\citet{garcia2017cliel} design a system named CLIEL for extracting core information from commercial law documents. Specifically, CLIEL identifies five types of contract elements using rule-based layout detection. \citet{azzopardi2016integrating} propose a hybrid approach based on regular expressions and provide a contract editing tool for lawyers. 

Recent \ac{CEE} methods are developed with deep learning and formulate the \ac{CEE} task as sequence labeling. 
\citet{chalkidis2017deep} explore deep learning methods for the \ac{CEE} task and employ a BiLSTM without manually written rules. 
\citet{sun2019toi} define seven semantic-specific clause categories and introduce a TOI pooling layer for the nested elements. 
\citet{chalkidis2019neural} revisit the \ac{CEE} task and explore how sequence encoders, CRF layers, and input representations affect the extractors. 

Existing \ac{CEE} approaches cannot be applied to the cross-domain scenario because of differences in context genres and element types.

\subsection{Cross-domain named entity recognition}
The cross-domain \ac{NER} task aims to identify named entities in a target domain using the shared knowledge from a source domain.
Recently, methods based on deep neural networks have been proposed for cross-domain \ac{NER}. 
\citet{pan2013transfer} design transfer joint embeddings for this task.
\citet{DBLP:conf/emnlp/QuFZHB16} model the correlation between source and target entity types with a two-layer neural network. 
To investigate the transferability of model components, \citet{DBLP:conf/iclr/YangSC17} present three sharing architectures for different transfer learning scenarios. 
Moreover, \citet{DBLP:conf/acl/JiaXZ19} propose a parameter generation network and incorporate the language modeling (LM) task to deal with zero-shot learning settings.
Unlike the above parameter-sharing frameworks, parameter transfer approaches first initialize the target model on the source domain \ac{NER} or LM~\citep{DBLP:conf/lrec/LeeDS18, sachan2018effective}, and then fine-tune the original model on the labeled \ac{NER} data from the target domain. 
\citet{DBLP:conf/emnlp/LinL18} add three neural adaptation layers (word adaptation, sentence adaptation, and output adaptation) to an existing \ac{NER} method. 
\citet{DBLP:conf/acl/JiaZ20} transfer entity type level knowledge using a multi-cell compositional LSTM structure and model each entity type using a separate cell state. For a thorough review of other techniques,  please refer to~\citep{li2020survey}.

We mainly focus on cross-domain \acf{CEE}. To the best of our knowledge, ours is the first study to concentrate on this task. The most closely related task is the cross-domain \ac{NER} task. However, cross-domain contract element extractors need to address adaptations of larger extraction zones and more fine-grained element types.
In our proposed model \ac{Bi-FLEET}, to handle the transfer of fine-grained element types, invariant relations between clauses and elements are captured by constructing a \ac{C-E} graph across domains (see below). In addition, a multi-task framework with a bi-directional feedback scheme is designed to reduce the impact of more diverse contexts in larger extraction zones. 

%% file: sections/method.tex

\section{Method}
\label{sec:method}
\begin{figure}[t]
    \centering
    \includegraphics[width=0.9\linewidth]{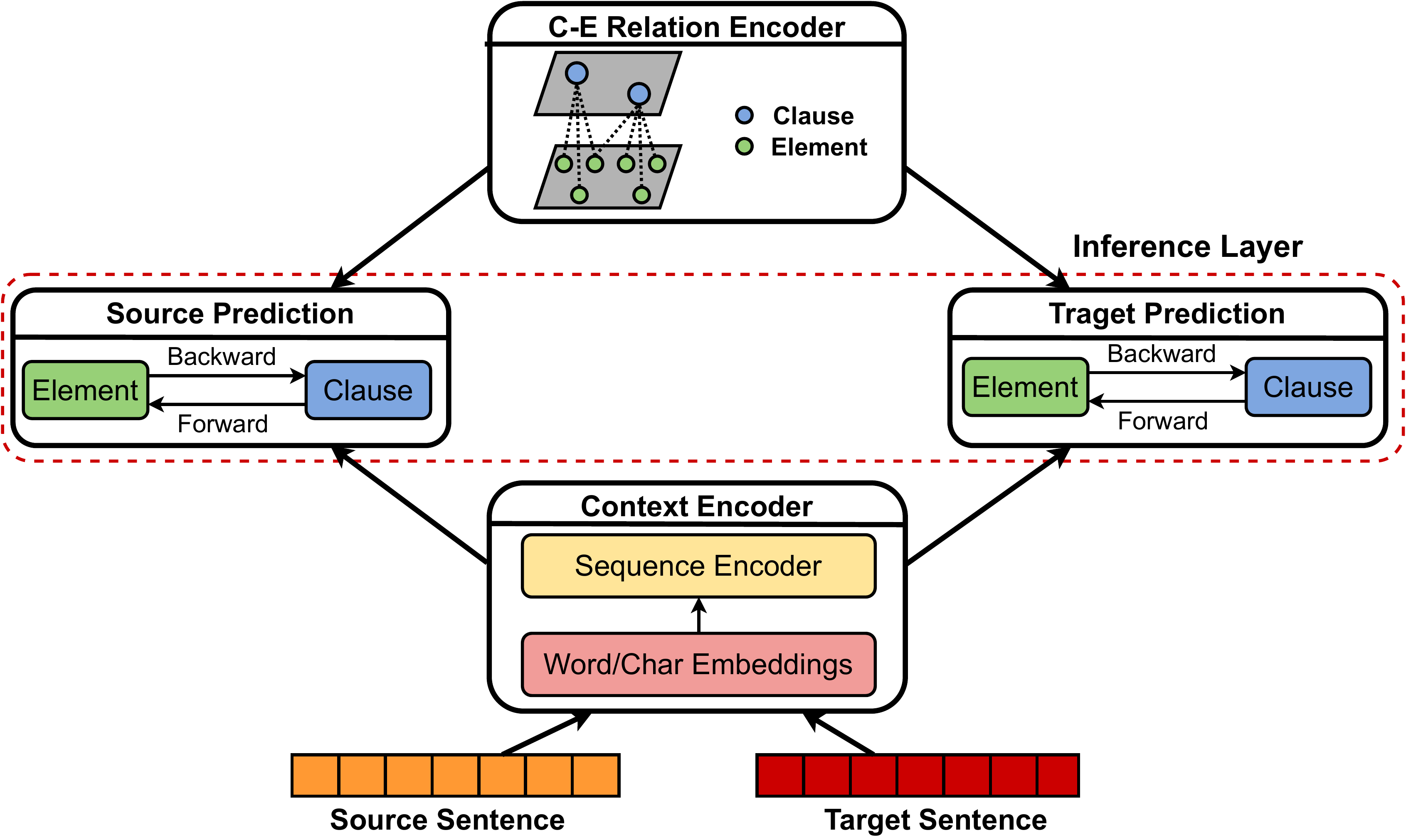}
    \caption{An overview of \ac{Bi-FLEET}. \ac{Bi-FLEET} has three components: a context encoder, a clause-element relation encoder, and an inference layer.}
    \Description{The framework of our model.}
    \label{fig:framework}
\end{figure}

In this section, we describe the \ac{Bi-FLEET} framework, our proposed method for \ac{CEE}. 
As illustrated in Fig.~\ref{fig:framework}, the framework has three main components: a context encoder, a clause-element relation encoder, and an inference layer. 
The context encoder, which includes an input embedding layer and sequence encoder, embeds every word in a sentence from a given clause. 
The clause-element relation encoder is shared by the source and target domain and calculates representations of clause and element types. 
The word embeddings and type representations are input to the inference layer for clause classification and contract element extraction across domains.

Next, we formulate the cross-domain \ac{CEE} task and compare it to the cross-domain \ac{NER} task (Sec.~\ref{subsec:problem formulation}). 
Then, we explain \ac{Bi-FLEET}'s context encoder (Sec.~\ref{subsec:context encoder}), the clause-element relation encoder (Sec.~\ref{subsec:C-E graph}), and the inference layer (Sec.~\ref{subsec:inference layer}). 
Finally, the loss functions and training process (Sec.~\ref{subsec:traning objectives}) are presented. 

\smallnegskip
\subsection{Problem formulation}
\label{subsec:problem formulation}
We write $\mathbf{C} = (\mathbf{s}_{1}, \mathbf{s}_{2}, \ldots, \mathbf{s}_{n})$ for  a clause from a contract, where $\mathbf{s}_{i}$ is the $i$-th sentence in the clause $\mathbf{C}$; $\mathbf{s}_{i} = (x_{i,1}, x_{i, 2},\ldots, x_{i, m})$, where $x_{i, j}$ is the $j$-th word in sentence $\mathbf{s}_{i}$. A \emph{contract element} $\mathbf{e}$ in the clause $\mathbf{C}$ is a sequence of words in one sentence: $\mathbf{e} = \{(x_{i,start}, x_{i, start+1}$, \ldots, $x_{i, end}), l^{e}\}$, where $l^{e}$ is the type label of the element $\mathbf{e}$ (such as payment period or deposit rate). The aim of the \acfi{CEE} task is to find the element $\mathbf{e}$ in the clause $\mathbf{C}$.

For the cross-domain \ac{CEE} task, there are $N_{s}$ labeled clauses in the source domain $\mathcal{S}$, as well as $N_{t}$ labeled clauses and $N_{u}$  unlabeled clauses in the target domain $\mathcal{T}$. 
Our goal is to transfer the contract element extractor to the target domain from the source domain. Specifically, the extractor is trained on the $N_{s}$ labeled clauses in the source domain and $N_{t}$ labeled clauses in the target domain to detect all the elements for the $N_{u}$ unlabeled clauses in the target domain. 

A similar task is cross-domain \acf{NER}, which focuses on identifying several types of named entities in only one sentence. In contrast, cross-domain \ac{CEE} aims at extracting much more fine-grained elements in multiple sentences. As mentioned before, the transfer of larger extraction zones and more element categories brings new challenges to cross-domain \ac{CEE}.

\smallnegskip
\subsection{Context encoder}
\label{subsec:context encoder}
Similar to previous \ac{CEE} and \ac{NER} methods~\cite{chalkidis2017deep, chalkidis2019neural, DBLP:conf/iclr/YangSC17, DBLP:conf/acl/JiaZ20}, our context encoder vectorizes words in a sentence from a clause by using two components, an input embedding layer and a sequence encoder.

\subsubsection{Input embedding layer}
Given an input sentence $\mathbf{s} = \{x_{1}, x_{2}$, $x_{3}, \ldots, x_{m} \} $ from a clause $\mathbf{C}$, to capture the word-level and character-level features, each word $x_{i}$ in the sentence $\mathbf{s}$ is embedded as the concatenation of the word embedding and the output of a char-level CNN:
\begin{equation}
    \small
    \mathbf{v}_{i} = \mathbf{E}^{w}(x_{i}) \oplus {\rm CNN}(\mathbf{E}^{c}(x_{i})),  
\end{equation}
where $\mathbf{E}^{w}$ denotes a shared word-level embedding lookup table, and $\mathbf{E}^{c}$ denotes a shared char-level embedding lookup table. $\rm{CNN}(\cdot)$ is a standard convolutional neural network operating on the character-level embedding sequence $\mathbf{E}^{c}(x_{i})$ for word $x_{i}$ in sentence $\mathbf{s}$. And $\oplus$ denotes the  vector concatenation operator.

\subsubsection{Sequence encoder}
To encode sentence-level features, a standard bi-directional LSTM layer~\cite{graves2005framewise} is adopted here. Given an input embedding sequence $\mathbf{v} = [\mathbf{v}_{1}, \mathbf{v}_{2}, \ldots, \mathbf{v}_{m}]$, at each time step, the Bi-LSTM layer calculates the current hidden vectors based on memory cells. The hidden outputs of the task and domain-specific Bi-LSTM unit can be written as follows:
\begin{equation}
    \small
    \label{eq:bi-lstm}
        \overrightarrow{\mathbf{h}}^{d,t}_{i} = {\rm LSTM}(\overrightarrow{\mathbf{h}}^{d,t}_{i-1}, \mathbf{v}_{i}),\quad
        \overleftarrow{\mathbf{h}}^{d,t}_{i} = {\rm LSTM}(\overleftarrow{\mathbf{h}}^{d,t}_{i+1}, \mathbf{v}_{i}),
\end{equation}
where $\mathbf{h}^{d,t}_{i}$ denotes the output hidden vector for domain $d\in \{\mathcal{S}, \mathcal{T}\}$  and task $t\in \{CC, CEE\}$ (clause classification and contract element extraction) with the input word $x_{i}$; $\rightarrow$ and $\leftarrow$ represent the directions of the LSTM unit. 
Note that a variety of methods can be adopted as the context encoder. In our experiments, BERT~\cite{devlin2018bert} is adopted as the input embedding layer to generate the contextualized word embeddings. Besides, the sequence encoders based on  the parameter generation network~\cite{DBLP:conf/acl/JiaXZ19} and multi-cell compositional LSTM~\cite{DBLP:conf/acl/JiaZ20} are evaluated as well. 

\subsection{Clause-element relation encoder}
\label{subsec:C-E graph}
To model relations between element and clause classes, we establish the \acf{C-E} graph and learn representations of the clause and element types with a GNN-based \ac{C-E} relation encoder. 

\subsubsection{C-E graph constructions}
\begin{figure}[t] 
    \centering
    \includegraphics[width=0.8\linewidth]{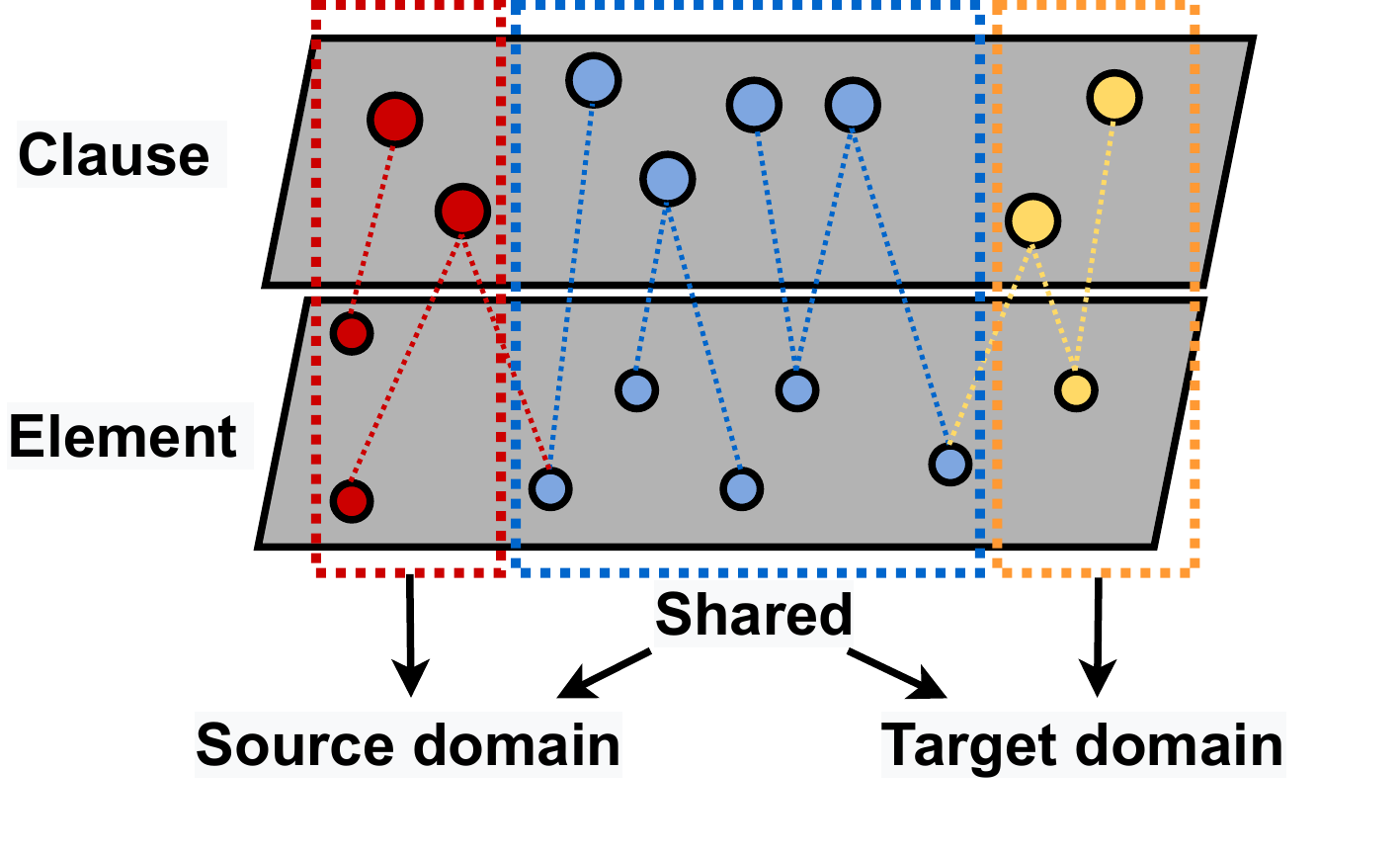}
    \caption{\ac{C-E} graph construction. Blue nodes are the shared element and clause types, while red and yellow ones are the ``not shared'' types between the source and target domains.}
    \Description{C-E graph constructions.}
    \label{fig:constructions}
\end{figure}

Before encoding clause and element types with a hierarchical graph neural network, we construct the \ac{C-E} graph.
As mentioned above, contract elements are strongly connected with categories of clauses.
For a clause type $ct$ and element type $et$ in a given domain $d$, to identify the relation between them,  we first define the probability $p_{et, ct}$, that the elements of type $et$ exist in clauses of type $ct$, and $p_{ct, et}$, that clauses of type $ct$ contain elements of type $et$, as follows: 
\begin{equation}
    \small
    \label{eq:constructions}
        p_{et, ct} = f_{et,ct}/ f_{et},\quad p_{ct, et}=f_{et,ct} / f_{ct},
\end{equation}
where, in a given domain $d$, $f_{et,ct}$ is the frequency that elements of type $et$ and clauses of type $ct$ co-occur. Here, $f_{et}$ and $f_{ct}$ are the frequencies that the elements of  type $et$ and clauses of type $ct$ appear in the given domain, respectively. Note that we only count once when multiple elements of the same types exist in one clause. When both $p_{et, ct}$ and $p_{ct, et}$ are larger than the preset threshold $\theta$, the element type $et$ and clause type $ct$ are connected in the C-E graph of the given domain $d$. 
As Fig.~\ref{fig:constructions} illustrates, we first select the edges between clause and element types (red and blue nodes) to establish the \ac{C-E} graph of the source domain. And then, for the target domain, we only maintain the shared types (blue nodes) and add the domain-specific types (yellow nodes) into the original graph.

\subsubsection{The GNN-based relation encoder} Given a \ac{C-E} graph $G$ in the domain $d$, we apply a hierarchical graph neural network to encode the element and clause types. Recent GNN-based frameworks follow the neighbor aggregation strategy~\cite{DBLP:conf/iclr/XuHLJ19,wang2019robust,feng2021causalgcn,feng2019graph}, updating the embeddings of the central node by iteratively aggregating the information of its neighbors. 
The $l$-th layer of a GNN, composed of an aggregation function ${\rm AGGREGATE(\cdot)}$ and a combination function ${\rm COMBINE(\cdot)}$, can be presented as follows:
\begin{equation}
    \small
    \begin{split}
        a_{v}^{(l)} &= {\rm AGGREGATE}^{(l)}\left(h_{u}^{(l-1)}:u \in \mathcal{N}_v\right) \\
        h_{v}^{(l)} &= {\rm COMBINE}^{(l)}\left(h_{v}^{(l-1)}, a_{v}^{(l)}\right),   
    \end{split}
\end{equation}
where $h_{v}^{l}$ is the feature vector of node $v$ at the $l$-th GNN layer and $\mathcal{N}_v$ is the set of nodes adjacent to the central node $v$.
We set two transformation matrices $\mathbf{W}_{c\rightarrow e}$ and $\mathbf{W}_{e\rightarrow c}$ for the clause-to-element and element-to-clause relations respectively, and the $l$-th layer of our proposed C-E relation encoder can be formulated:
\vspace{-1mm}
\begin{equation}
    \label{eq:relation encoder}
    \small
    \begin{split}
       \mathbf{ele}_{p}^{d, (l)}=\sigma\left(\sum_{u\in \mathcal{N}^{d,e}_{p}}{\alpha_{u}^{d, c}\mathbf{W}_{c\rightarrow e}\cdot\mathbf{cla}_{u}^{d, (l-1)}}+\mathbf{W}_{e}\cdot\mathbf{ele}_{p}^{d, (l-1)}\right)  \\
       \mathbf{cla}_{k}^{d, (l)}=\sigma\left(\sum_{v\in \mathcal{N}^{d, c}_{k}}{\alpha_{v}^{d, e}\mathbf{W}_{e\rightarrow c}\cdot\mathbf{ele}_{v}^{d, (l-1)}}+\mathbf{W}_{c}\cdot\mathbf{cla}_{k}^{d,  (l-1)}\right),
    \end{split}
\end{equation}

\noindent where, given a domain $d$, $\mathbf{ele}_{p}$ and $\mathbf{cla}_{k}$ are the representations of the element type $et_{p}$ and clause type $ct_{k}$ separately, while $\alpha_{v}^{e}$ and $\alpha_{u}^{c}$ are the trainable weights for the $v$-th element type and $u$-th clause, type respectively. Given a \ac{C-E} graph of domain $d$, $\mathcal{N}^{e}_{p}$ is the set of neighborhood clause types for $et_{p}$, and $\mathcal{N}^{c}_{k}$ is the set of neighborhood element types for $ct_{k}$. $\mathbf{W}_{e}$ and $\mathbf{W}_{c}$ denote the matrices for the  self-connection. 
$\sigma(\cdot)$ is the activation function.
$\mathbf{W}_{c\rightarrow e}$, $\mathbf{W}_{e\rightarrow c}$, $\mathbf{W}_{e}$ and $\mathbf{W}_{c}$ are shared in both source and target domains. 
If a clause or element types exists in both domains 
, their representations are also shared across domains. 

\subsection{Inference layer}
\label{subsec:inference layer}
Given the sentence-level feature and type representations of elements and clauses,  the multi-task inference layer aims to predict the clause category labels (\acf{CC}) and  extract contract elements from clauses (\acf{CEE}). To capture the interactions between the two tasks above, we design a novel bi-feedback scheme in the inference layer. In this section, we demonstrate the bi-feedback scheme first and then explain how to extract elements with forward information and verify the clause classification results using backward information. 
\begin{figure}[t] 
    \centering
    \includegraphics[width=0.85\linewidth]{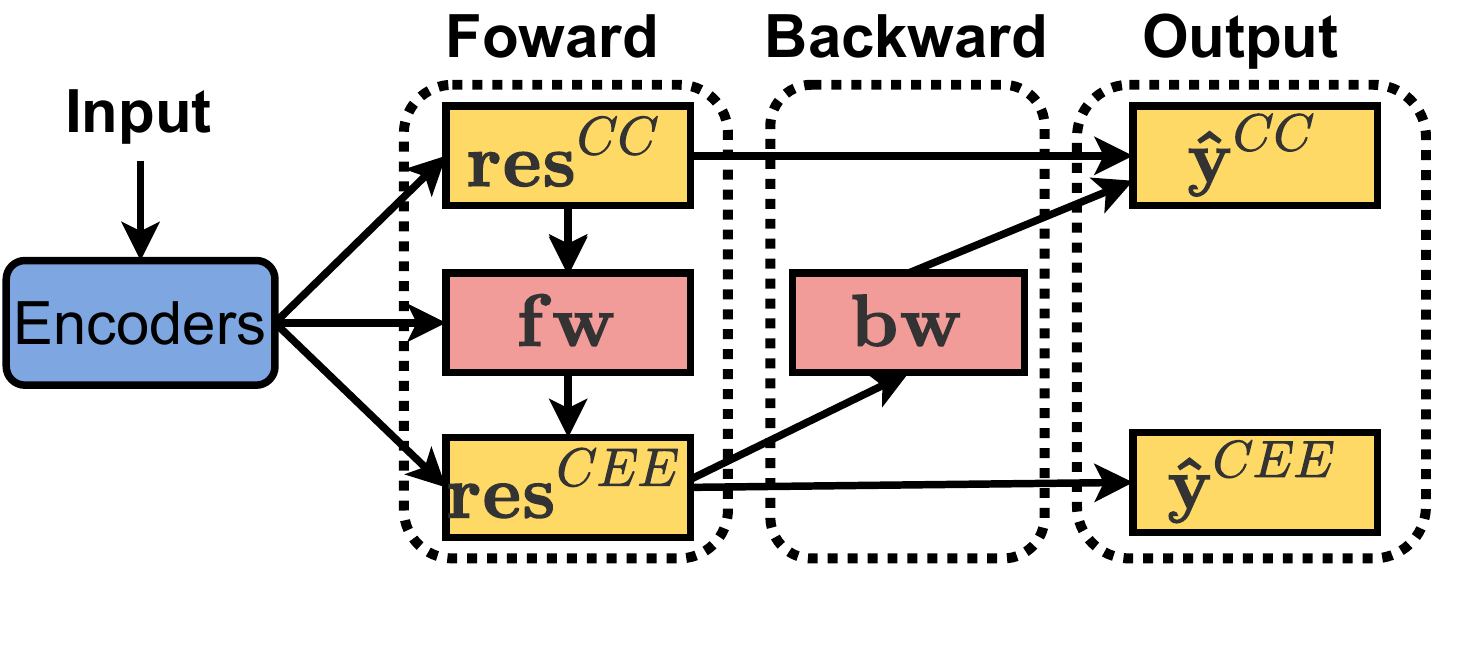}
    \caption{The inference layer with the bi-feedback scheme.  }
    \Description{bi-feedback scheme.}
    \label{fig:bifeedback}
\end{figure}

\subsubsection{The bi-feedback scheme}
Given a clause from the contract, a human annotator usually first decides the clause type and then identifies the elements in the clause. In that case, the classification of clauses often facilitates the element extraction process, while the identification of elements can also verify the clause type predictions. Based on this observation, as shown in Fig.~\ref{fig:bifeedback}, we design a multi-task framework with the bi-feedback scheme.
Given a training instance for task $t\in \{CC, CEE\}$, assuming that the result distribution vector $\mathbf{res}^{t}$ is predicted, we combine the result distribution vector with the clause or element representations , and obtain the latent type vector $\mathbf{ltv}^{t}$ for $\mathbf{res}^{t}$ as follows:
\begin{equation}
    \small
    \label{eq:latent class}
        \mathbf{ltv}^{CC} = \sum_{k}{res^{CC}_{k}\cdot \mathbf{cla}_{k}},\quad
        \mathbf{ltv}^{CEE} = \sum_{p}{res^{CEE}_{p}\cdot \mathbf{ele}_{p}}, 
\end{equation}
where, given a task $t$, $res_{k}$  denotes the $k$-th type of the results. $\mathbf{cla}_{k}$ and $\mathbf{ele}_{p}$ are the representations of the $k$-th clause type or $p$-th element type. 
Then, we calculate the forward information $\mathbf{fw}$ and backward information $\mathbf{bw}$ based on the latent type vector $\mathbf{ltv}^{t}$:
\begin{equation}
    \centering
    \small
    \label{eq:forward and backward}
        \mathbf{fw} =\sigma(\mathbf{W}_{f}\cdot \mathbf{ltv}^{CC}),\quad
        \mathbf{bw} =\sigma(\mathbf{W}_{b}\cdot \mathbf{ltv}^{CEE}),
\end{equation}
where $\mathbf{W}_{f}$ and $\mathbf{W}_{b}$ are the transformation matrices. Note that we set different transformation matrices and biases for source and target domains. In the rest of this section, we explain how to predict the result distribution $\mathbf{res}^{t}$ and obtain the final predictions, $\mathbf{\hat{y}}^{CC}$ and $\mathbf{\hat{y}}^{CEE}$, using forward and backward information. In our multi-task framework, $\mathbf{fw}$ and $\mathbf{bw}$ serve as gates to filter out the noise and reduce prediction mistakes.

\subsubsection{Clause classification (CC)}
To classify the clause that the given sentence $\mathbf{s}$ belongs to, given a domain $d$, we use the sentence-level feature $\mathbf{h}=[\overrightarrow{\mathbf{h}}_{1}\oplus\overleftarrow{\mathbf{h}}_{1}, \overrightarrow{\mathbf{h}}_{2}\oplus\overleftarrow{\mathbf{h}}_{2}, \ldots ,\overrightarrow{\mathbf{h}}_{m}\oplus\overleftarrow{\mathbf{h}}_{m}]$ from the Bi-LSTM. The $k$-th element of the result distribution vector $\mathbf{res}^{CC}$ is predicted by  
\begin{equation}
    \centering
    \small
    \label{eq: res CC}
    res^{CC}_{k}=\sigma(\mathbf{w}_{CC}\cdot[\mathbf{\hat{h}}\oplus \mathbf{cla}_{k}] + b_{CC}),
\end{equation}
where $\mathbf{\hat{h}} = \sum_{i}{\alpha_{i}\cdot \mathbf{h}_{i}}$ and $\sum_{i}{\alpha_{i}}=1$. $\mathbf{\hat{h}}$ represents the overall feature of the given sentence and $\alpha_{i}$ denotes the trainable attention weight for word $x_{i}$. Considering the different label sets across domains, the parameters, $\mathbf{w}_{CC}$ and $b_{CC}$, are not shared. Based on the result distribution vector $\mathbf{res}^{CC}$, forward information $\mathbf{fw}_{s}$ for sentence $\mathbf{s}$ can be calculated by Eq.~\ref{eq:latent class} and~\ref{eq:forward and backward}.
The final prediction $\mathbf{\hat{y}}^{CC}$ can be verified by the backward information as follows:
\begin{equation}
    \label{eq:y CC}
    \small
    \mathbf{\hat{y}}^{CC}= {\rm softmax}(\mathbf{bw}_{1}\otimes\mathbf{bw}_{2}\otimes \cdots \otimes\mathbf{bw}_{m}\otimes \mathbf{res}^{CC}),
\end{equation}
where $\mathbf{bw}_{i}$ denotes the backward information for word $x_{i}$ in sentence $\mathbf{s}$ (from the contract element extraction task). ${\rm softmax}(\cdot)$ is the softmax function. $\otimes$ is the element-wise product operator.

\subsubsection{The contract element extraction}
Similar to the \ac{NER} task, standard CRFs~\citep{DBLP:conf/acl/MaH16} are adopted in the inference layer for \ac{CEE}.
Given a sentence $\mathbf{s}$ from the contract clause, the sentence-level feature $\mathbf{h}=[\overrightarrow{\mathbf{h}}_{1}\oplus\overleftarrow{\mathbf{h}}_{1}, \overrightarrow{\mathbf{h}}_{2}\oplus\overleftarrow{\mathbf{h}}_{2}, \ldots ,\overrightarrow{\mathbf{h}}_{m}\oplus\overleftarrow{\mathbf{h}}_{m}]$, where $\mathbf{h}_{i}=\overrightarrow{\mathbf{h}}_{i}\oplus\overleftarrow{\mathbf{h}}_{i}$, can be obtained from the context encoder. We merge $\mathbf{fw}_{s}$ (from the clause classification task for sentence $\mathbf{s}$) with each hidden vector in $\mathbf{h}$:
\begin{equation}
    \small
    \label{eq:h FD}
    \mbox{}\hspace*{-1mm}
    \begin{split}
    &\mathbf{h}^{fw}= {}\\
    &[{\rm norm}(\mathbf{h}_{1}\otimes\mathbf{fw}_{s}), {\rm norm}(\mathbf{h}_{2}\otimes\mathbf{fw}_{s}), \ldots ,{\rm norm}(\mathbf{h}_{m}\otimes\mathbf{fw}_{s})],   
    \hspace*{-1mm}\mbox{}
    \end{split}
\end{equation}
where ${\rm norm}(\cdot)$ is the normalization function. $\otimes$ is the element-wise product operator. Then, the output probability $p(\mathbf{\hat{y}}^{CEE}|\mathbf{s})$ of the final prediction sequence $\mathbf{\hat{y}}^{CEE} = [l_{1}, l_{2}, \ldots, l_{m}]$ over the input sentence $\mathbf{s}$ can be calculated by:
\begin{equation}
    \small
    \label{eq: CRF}
    p(\mathbf{\hat{y}}^{CEE}|\mathbf{s})=\frac{{\rm exp}\{\sum_{i}(\mathbf{w}_{CRF}^{l_{i}}\cdot \mathbf{h}^{fw}_{i}+b_{CRF}^{{l}_{i-1}, l_{i}})\}}{\sum_{\mathbf{\hat{y}}'}{\rm exp}\{\sum_{i}(\mathbf{w}_{CRF}^{l'_{i}}\cdot \mathbf{h}^{fw}_{i}+b_{CRF}^{l'_{i-1}, l'_{i}})\}},
\end{equation}
where $\mathbf{\hat{y}}'$ denotes an arbitary label sequence; $\mathbf{w}_{CRF}^{l_{i}}$ is the model parameter for $l_{i}$, and $b_{CRF}^{l_{i-1}, l_{i}}$ is the bias for $l_{i-1}$ and $l_{i}$. As Fig.~\ref{fig:framework} shows, we use CRF($s$) and CRF($t$) for the different label sets in the source and target domains respectively. 

To effectively search the highest scored label sequence, the first-order Viterbi algorithm is used. We consider the result distribution vector $\mathbf{res}^{CEE}_{i}$ for the word $x_{i}$ in the given sentence $s$ as a one-hot vector.
$res^{CEE}_{i, p} = 1$ only when the word $x_{i}$ is classified as the $p$ th element type. Then, the backward information $\mathbf{bw}_{i}$ for word $x_{i}$ is obtained by Eq.~\ref{eq:latent class} and~\ref{eq:forward and backward}.

\subsection{Joint training}
\label{subsec:traning objectives}
Given the labeled dataset $\mathcal{D}=\{(\mathbf{s}_{n},\mathbf{y}^{CC}_{n},\mathbf{y}^{CEE}_{n})\}_{n=1}^{N}$, loss functions for the clause classification task ($\mathcal{L}_{CC}$) and the contract element extraction task ($\mathcal{L}_{CEE}$) are defined over the training dataset.
Specifically, for the clause classification task, the cross-entropy loss function is used as follows:
\begin{equation}
    \small
    \label{eq: loss CC}
    \mathcal{L}_{CC} = -\frac{1}{N}\sum_{n=1}^{N}{\mathbf{y}^{CC}_{n}{\rm log}(p(\mathbf{\hat{y}}^{CC}_{n}))}.
\end{equation}
For the contract element extraction task, the sentence-level negative log-likelihood loss is adopted for the training process:
\begin{equation}
    \small
    \label{eq: loss CEE}
    \mathcal{L}_{CEE} = -\frac{1}{N}\sum_{n=1}^{N}{{\rm log}(p(\mathbf{y}^{CEE}_{n}|\mathbf{s}_{n}))}.
\end{equation}
And the overall loss is defined as follows:
\begin{equation}
    \small
    \label{eq: loss}
    \mathcal{L} = \sum_{d\in\{\mathcal{S}, \mathcal{T}\}}{\lambda^{d}(\mathcal{L}^{d}_{CEE}+\lambda^{t}\mathcal{L}^{d}_{CC})} + \frac{\lambda}{2}\|\Theta \|^{2},
\end{equation}
where $\lambda^{d}$ is the domain weight and $\lambda^{t}$ is the task weight. $\lambda$ is the regularization weight for the parameter set $\Theta$.

As Algorithm~\ref{algorithm:multi-task training} shows, we design a cross-domain and multi-task training process. Before training, the C-E graphs are constructed (line 1). In each training time step, we encode the sentence-level feature $\mathbf{H}^{d,t}$ (line 4) and type representations, $\mathbf{ele}^{d}$ and  $\mathbf{cla}^{d}$ (line 5). Then, the result distribution $\mathbf{res}^{d, CC}$ for clause classification is predicted (line 6) and forward information $\mathbf{fw}^{d}$ is computed based on the result distribution (line 7). For the contract element extraction, the element labels $\mathbf{\hat{y}}^{CEE}$ are calculated with forward information (line 8). Next, the backward information $\mathbf{bw}^{d}$ is computed (line 9) and the final predictions $\mathbf{\hat{y}}^{CC}$ for clause classification are verified (line 10). Finally,  the overall loss function $\mathcal{L}$ is obtained (line 11) and the model parameters are jointly updated (line 13).  
\begin{algorithm}[t]
    \caption{Joint Training.}
    \label{algorithm:multi-task training}      
    \begin{algorithmic}[1] 
    \Require Training data $\mathcal{D}_{s}$ and $\mathcal{D}_{t}$; parameter sets for the context encoder $\theta_{cont}$, C-E relation encoder $\theta_{C-E}$ (including $\mathbf{W}_{c\rightarrow e}$, $\mathbf{W}_{e\rightarrow c}$, $\mathbf{W}_{c}$, $\mathbf{W}_{e}$, $\alpha^{e}$, and $\alpha^{c}$), and inference layer $\theta_\mathit{infer}$ (including $\mathbf{W}_{f}$, $\mathbf{W}_{b}$, $\mathbf{w}_{CC}$, $b_{CC}$, $\mathbf{w}_{CRF}$, and $b_{CRF}$).
    \Ensure Target-domain model;
    \State Contruct the C-E graphs $G^{\mathcal{S}}$ and $G^{\mathcal{T}}$ for both domains (Eq.~\ref{eq:constructions});
    \While{Training process not terminated}
    \For{$d$ in $\{\mathcal{S}, \mathcal{T}\}$}
    \State Compute sentence-level feature $\mathbf{h}^{d,t}\leftarrow {\rm BiLSTM(\mathbf{v})}$ 
    
    \mbox{}\hfill(Eq.~\ref{eq:bi-lstm});
    \State Calculate class representations $\mathbf{ele}^{d}$, $\mathbf{cla}^{d}$ (Eq.~\ref{eq:relation encoder});
    \State Predict clause labels $\mathbf{res}^{d, CC}$ with $\mathbf{h}^{d,CC}$ and $\mathbf{cla}^{d}$ (Eq.~\ref{eq: res CC});
    \State Compute the forward information $\mathbf{fw}^{d}$ with $\mathbf{res}^{d, CC}$ 
    
    \mbox{}\hfill(Eq.~\ref{eq:forward and backward});
    \State Predict element labels $\mathbf{\hat{y}}^{CEE}$ with $\mathbf{h}^{d,CEE}$ and $\mathbf{fw}^{d}$; 
    
    \mbox{}\hfill(Eq.~\ref{eq:h FD}--\ref{eq: CRF});
    \State Compute the backward information $\mathbf{bw}^{d}$ with $\mathbf{res}^{d, CEE}$ 
    
    \mbox{}\hfill(Eq.~\ref{eq:forward and backward});
    \State Verify the final prediction $\mathbf{\hat{y}}^{CC}$ using $\mathbf{bw}^{d}$ (Eq.~\ref{eq:y CC});
    \State Compute the overall loss function $\mathcal{L}$ (Eq.~\ref{eq: loss CC}--\ref{eq: loss});
    \EndFor
 \State Update $\theta_{cont}$, $\theta_{C-E}$, and $\theta_\mathit{infer}$ based on $\mathcal{L}$.
 \EndWhile
    \end{algorithmic}
    \end{algorithm}

%% file: sections/experiments.tex

\section{Experiments}
\label{sec:experiments}
\subsection{Research questions}
We aim to answer the following research questions:
\begin{enumerate*}[label=(RQ\arabic*),leftmargin=*,nosep]
\item Does \ac{Bi-FLEET} outperform state-of-the-art methods on the cross-domain CEE taks? (Sec.~\ref{subsection:CEE})
\item Can \ac{Bi-FLEET} be generalized to the cross-domain NER task? (Sec.~\ref{subsection:NER})
\end{enumerate*}
 
\subsection{Datasets}
\label{subsec:datasets}
\begin{table}
  \small
  \centering
  \caption{Statistics of the CEE datasets.}
  \label{tab:CEE dataset}
  \begin{tabular}{lccll}
  \toprule
  \multirow{2}*{Dataset} & \multicolumn{2}{c}{Type} & \multirow{2}*{Size} &\multirow{2}*{Train./Valid./Test} \\
  \cmidrule{2-3}
   ~ & Element & Clause & ~ & ~ \\
  \midrule
    \multirow{3}*{Individual} & \multirow{3}*{70} & \multirow{3}*{18} & \#Pos Sentence & 13.6K/1.7K/1.7K \\
    ~ & ~ & ~ & \#Neg Sentence& 60.0K/7.5K/7.5K \\
    ~ & ~ & ~ & \#Element& 25.7K/3.3K/3.3K \\
  \midrule
  \multirow{3}*{Commercial} & \multirow{3}*{79} & \multirow{3}*{17} & \#Pos Sentence & 4.8K/0.6K/0.6K \\
  ~ & ~ & ~ & \#Neg Sentence& 19.7K/2.5K/2.5K \\
  ~ & ~ & ~ & \#Element& 8.7K/1.2K/1.1K \\
  \bottomrule
\end{tabular}
\end{table}

\begin{table}
  \small
  \centering
  \caption{Statistics of the \ac{NER} datasets.}
  \label{tab:NER dataset}
  \begin{tabular}{l lll}
  \toprule
    Dataset & Entity Type& Size & Train./Valid./Test \\
  \midrule
    \multirow{2}*{BioNLP13PC} & CHEM,CC, &\#Sentence& 2.5K/0.9K/1.7K \\
    ~ & GGP& \#Entity& 7.9K/2.7K/5.3K\\
  \midrule
  \multirow{2}*{BioNLP13CG} & CHEM,CC, &\#Sentence& 3.0K/1.0K/1.9K \\
  ~ & GGP, etc.& \#Entity& 10.8K/3.6K/6.9K\\
  \midrule
  \multirow{2}*{CoNLL-2003} & PER, LOC, &\#Sentence& 15.0K/3.5K/3.7K \\
  ~ & ORG, MISC& \#Entity& 23.5K/5.9K/5.6K\\
  \midrule
  \multirow{2}*{Broad Twitter} & PER, LOC, &\#Sentence& 6.3K/1.0K/2.0K \\
  ~ & ORG & \#Entity& 8.8K/1.7K/4.4K\\
  \midrule
  \multirow{2}*{Twitter} & PER, LOC, &\#Sentence& 4.3K/1.4K/1.5K \\
  ~ & ORG, MISC& \#Entity& 7.5K/2.5K/2.5K\\
  \bottomrule
\end{tabular}
\end{table}

For cross-domain \ac{CEE}, we collect 340 open individual Chinese contracts from the web and 1,422 business Chinese contracts from partners. For every contract, at least two annotators with a legal background were asked to label the clauses and contract elements. The whole annotation process took about 3 months. To establish the datasets, we first split the contracts sentence by sentence, and filtered out elements that appear no more than 20 times and sentences with more than 100 characters. Then, these sentences were divided into training/validation/test sets with a 8/1/1 ratio, ensuring that all the elements and clauses in the validation and test splits occur at least once in the training set. Detailed statistics are provided in Table~\ref{tab:CEE dataset}. Different from NER, some sentences in clauses may not contain any elements. Such sentences are named ``Neg'' sentences in our experiments, while sentences in which elements do exist are called ``Pos'' sentences. 

For cross-domain \ac{NER}, we use five public datasets, BioNLP13PC and BioNLP13CG~\citep{DBLP:conf/bionlp/NedellecBKKOPZ13}, CoNLL-2003 English dataset~\citep{DBLP:conf/conll/SangM03}, Broad Twitter dataset~\citep{DBLP:conf/coling/DerczynskiBR16}, and Twitter dataset~\citep{DBLP:conf/acl/JiZCLN18}. Detailed statistics of these datasets are shown in Table~\ref{tab:NER dataset}. 

Following \citet{DBLP:conf/acl/JiaZ20}, we reconstruct two cross-domain \ac{CEE} datasets (I2C and C2I) and three cross-domain \ac{NER} datasets (BioNLP, Broad Twitter, and Twitter) based on the seven datasets mentioned above. For the I2C (or C2I) dataset, Commercial (or Individual) is used as the target-domain dataset, and the other contract dataset is the source-domain dataset. For the BioNLP dataset, the target-domain dataset is BioNLP13CG, while the source-domain dataset is BioNLP13PC. For the Broad Twitter dataset, Broad Twitter is the target-domain and CoNLL-2003 as the source-domain dataset. For the Twitter dataset, Twitter is used as the target-domain dataset, while CoNLL-2003 is the source-domain dataset. 

\subsection{Baselines}
\label{subsec:baselines}
For cross-domain \ac{CEE} and \ac{NER}, we compare Bi-FLEET with recent baselines. These methods can be categorized into three groups: 
\begin{itemize}[leftmargin=*,nosep]
  \item Target-domain only methods, BILSTM and MULTI-CELL LSTM \cite{DBLP:conf/acl/JiaZ20}, are trained without source-domain information.
  \item For cross-domain settings, we conduct comparisons with the LSTM-based MULTI-TASK (LSTM)~\citep{DBLP:conf/iclr/YangSC17} and two variants, MULTI-TASK+PGN and the method of \citet{DBLP:conf/acl/JiaXZ19}, both using parameter generation networks (PGN)  to generate parameters for source and target domain LSTMs. The multi-cell compositional LSTM based method of \citet{DBLP:conf/acl/JiaZ20} is considered too.
  \item In addition, we evaluate language model (LM) pretraining based methods, including BERT~\citep{devlin2018bert}, BIOBERT~\citep{DBLP:journals/bioinformatics/LeeYKKKSK20}, and the method of Jia and Zhang (BERT)~\citep{DBLP:conf/acl/JiaZ20} that leverages the outputs of LM pretraining methods as contextualized word embeddings.
\end{itemize}
As mentioned in Sec.~\ref{subsec:context encoder}, we adopt multiple types of context encoders. Specifically, \ac{Bi-FLEET} (LSTM) uses the standard LSTM layer as the sequence encoder. \ac{Bi-FLEET} (PGN) adds an PGN to generate parameters for domain-specific LSTMs. Similar to the MULTI-TASK (LSTM) and MULTI-TASK+PGN, both \ac{Bi-FLEET} (LSTM) and \ac{Bi-FLEET} (PGN) employ the max-margin principle~\citep{DBLP:conf/naacl/GimpelS10} for loss functions. \ac{Bi-FLEET} (MULTI-CELL) and \ac{Bi-FLEET} (BERT) utilize the multi-cell compositional LSTM and BERT as context encoders.

\subsection{Evaluation metrics}
Following~\citep{DBLP:conf/conll/SangM03,sun2019toi,DBLP:conf/acl/JiaXZ19, DBLP:conf/acl/JiaZ20}, we employ entity (or element) level Precision (P), Recall (R), F1-score (F1), and Accuracy for evaluation. Precision indicates the percentage of named entities (or elements) extracted by the method that are correct. Recall represents the percentage of entities (or elements) in the datasets that are predicted by the method. To this end, an entity (or element) is correct only when the corresponding entity (or element) in the dataset is exactly matched. The F1-score is the harmonic mean of Precision and Recall. We report P, R, and F1 for the cross-domain CEE task and F1 for the cross-domain NER task. We apply Accuracy to assess the proportion of clause type predictions that is correct.

\subsection{Implementation details}
\label{subsec:implementation details}
Our parameter settings mainly follow Jia and Zhang~\citep{DBLP:conf/acl/JiaZ20}. We use the NCRF++ toolkit~\citep{DBLP:conf/acl/YangZ18} to develop our models. All word embeddings are fine-tuned in the training process, while character embeddings are randomly initialized. The batch size is set to 30, and the initial learning rates for target-domain only, cross-domain, and BERT-based methods are 0.001, 0.0015, and 3e-5 respectively. The number of the layers of the \ac{C-E} relation encoder is set to 3.

For cross-domain CEE, word embeddings of 100 dimensions are pretrained on the Baidu Encyclopedia and Weibo corpus~\citep{DBLP:conf/emnlp/Cao000L18}. 
For BERT fine-tuning methods, we use the base-sized BERT~\citep{devlin2018bert} pretrained on the Chinese Wikipedia corpus. 

For cross-domain NER, word embeddings are initialized with PubMed 200 dimension vectors~\citep{DBLP:conf/bionlp/ChiuCKP16} for BioNLP experiments and GloVe 100 dimension vectors~\citep{DBLP:conf/emnlp/PenningtonSM14} for others.  In cross-domain NER, ``clause'' labels do not exist. In that case, we employ the base-sized BERT~\citep{devlin2018bert} and BIOBERT~\citep{DBLP:journals/bioinformatics/LeeYKKKSK20} models to embed each sentence and then cluster [CLS] vectors by the K-Means algorithm to generate the labels of sentences in an unsupervised manner. Specifically, we cluster sentences into 20 groups and represent these groups with the clause-element relation encoder.

%% file: sections/results.tex

\section{Experimental Results}
\label{sec:results}

\begin{figure}
  \centering
  \includegraphics[width=0.75\linewidth]{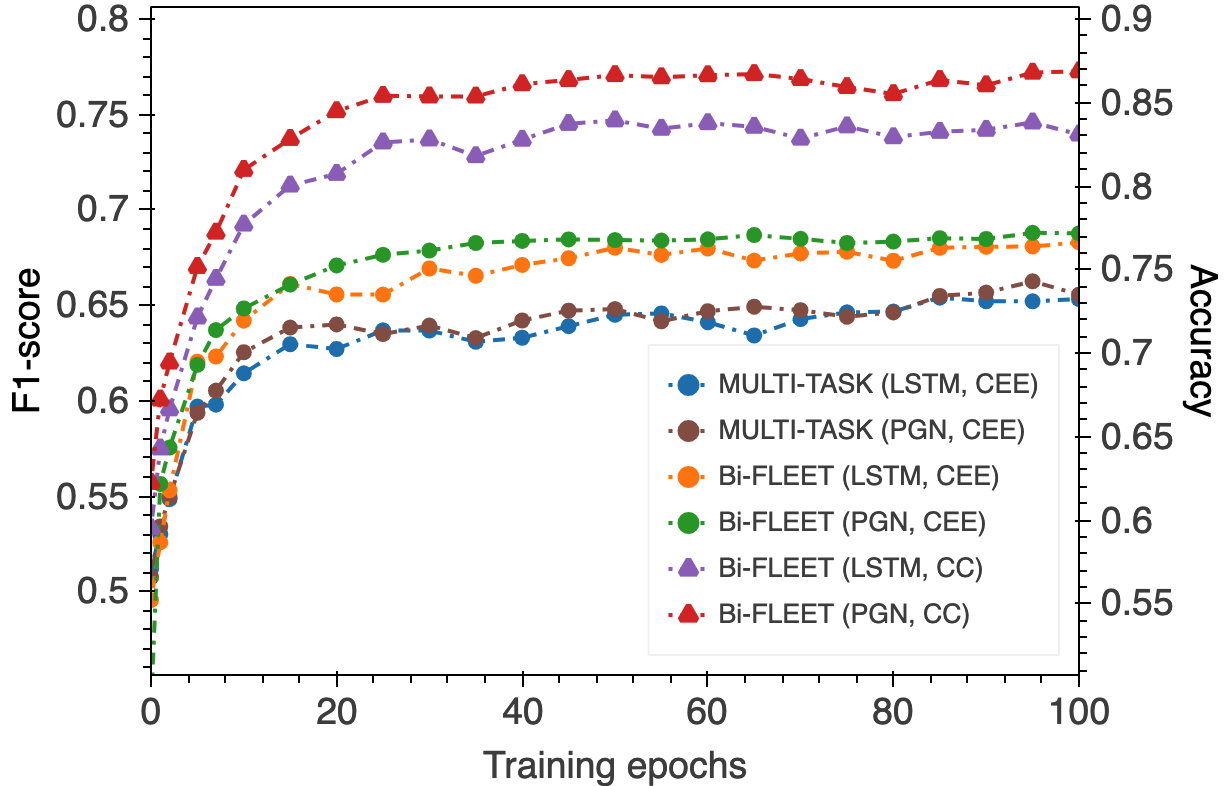}
  \caption{Learning curves on the I2C dataset.} 
  \label{fig:development}
\end{figure}

Prior to addressing our research questions RQ1 and RQ2, we examine the training process of Bi-FLEET. In particular, we examine whether cross-domain \acf{CEE} and \acf{CC} conflict with each other during the training process.
We use the I2C dataset for this purpose. In Fig.~\ref{fig:development}, the learning curves of cross-domain CEE and CC against the training epochs are shown. MULTI-TASK (LSTM) and MULTI-TASK+PGN are the cross-domain baselines. For comparison, we also show the learning curves of \ac{Bi-FLEET} (LSTM) and \ac{Bi-FLEET} (PGN). We see that:
\begin{itemize}[leftmargin=*,nosep]
  \item F1-scores of all models increase continuously before 60 training epochs, and reach a stable level when the number of training epoch rises to 100. 
  \item Cross-domain CEE and CC share the same improvement trend, and there is no conflict between the two tasks during the training process, indicating that both tasks can adopt the same form of feature space. 
\end{itemize}

\subsection{Cross-domain CEE}
\label{subsection:CEE}
\begin{table}[t]
  \small
  \setlength\tabcolsep{3.5pt}
  \caption{Results on two cross-domain CEE datasets. Significant improvements against the best baseline with the same context encoder are marked with $\ast$ (t-test, $p < 0.05$).}
  \label{tab:cross-domain CEE results}
  \begin{tabular}{l ccc ccc}
    \toprule
    \multirow{2}{*}{Model} & \multicolumn{3}{c}{I2C} & \multicolumn{3}{c}{C2I}\\
    \cmidrule(r){2-4}
    \cmidrule(r){5-7}
    ~ & P & R & F1 & P & R & F1 \\
  \midrule
    BILSTM (Random)  & 60.35  & 64.02 & 62.13  & 63.65  & 68.32 & 65.90 \\
    BILSTM  & 61.29  & 65.18 & 63.18  & 64.92  & 69.87 & 67.30 \\
    \midrule
    MULTI-TASK (Random) & 62.14 & 65.89 & 63.96 & 65.30 & 70.36 & 67.73 \\
    MULTI-TASK (LSTM)  & 63.21 & 67.50 & 65.28 & 65.86 & 71.61 & 68.62 \\
    MULTI-TASK+PGN  & 63.67 & 67.77 & 65.66 & 65.99 & 71.91 & 68.83 \\
    \citet{DBLP:conf/acl/JiaXZ19}  & 62.57 & 66.96 & 64.69 & 65.45 & 70.84 & 68.04 \\
    \midrule
    BERT-BASE  & 67.57 & 71.25 & 69.36 & 68.72 & 75.70 & 72.04 \\
    \midrule
    \ac{Bi-FLEET} (LSTM) & $66.07$\rlap{$^{\ast}$} & $70.20$\rlap{$^{\ast}$} & $68.07$\rlap{$^{\ast}$} & $67.98$\rlap{$^{\ast}$} & $75.26$\rlap{$^{\ast}$} & $71.44$\rlap{$^{\ast}$}\\ 
    \ac{Bi-FLEET} (PGN)  & $66.53$\rlap{$^{\ast}$} & $70.45$\rlap{$^{\ast}$} & $68.43$\rlap{$^{\ast}$} & $68.31$\rlap{$^{\ast}$} & $75.37$\rlap{$^{\ast}$} & $71.69$\rlap{$^{\ast}$} \\
    \ac{Bi-FLEET} (BERT) & $\mathbf{70.17}$\rlap{$^{\ast}$} & $\mathbf{73.75}$\rlap{$^{\ast}$} & $\mathbf{71.92}$\rlap{$^{\ast}$} & $\mathbf{70.71}$\rlap{$^{\ast}$} & $\mathbf{78.69}$\rlap{$^{\ast}$} & $\mathbf{74.49}$\rlap{$^{\ast}$} \\ 
  \bottomrule
\end{tabular}
\end{table}

We turn to RQ1. Table~\ref{tab:cross-domain CEE results} shows the experimental outcomes for the cross-domain CEE on the I2C and C2I datasets. To investigate the influence of pretrained word embeddings, we set up two more baseline models, BILSTM (Random) and MULTI-TASK (Random). In our experiments, the multi-cell compositional LSTM based models~\citep{DBLP:conf/acl/JiaZ20} overfit very easily, since their parameter counts increase drastically with element types; for this reason we do not report the results of multi-cell compositional LSTM based methods.
Based on Table~\ref{tab:cross-domain CEE results}, we can draw the following conclusions: 

\begin{itemize}[leftmargin=*,nosep]
  \item Cross-domain CEE is challenging, and for most baselines the F1-score is substantially less than 0.7. 
  In contrast, our proposed framework, Bi-FLEET, can effectively transfer contract element extractors from one domain to another. 
  \item The proposed model \ac{Bi-FLEET} significantly outperforms the baselines. Compared to the baseline models with the same context encoder, \ac{Bi-FLEET} (LSTM), \ac{Bi-FLEET} (PGN) and \ac{Bi-FLEET} (BERT) can obtain a consistent increase in P, R, and F1-scores, respectively. That is, integrated with the C-E relation encoder and bi-feedback scheme  improves the performance across the board.
  \item In comparison with target-domain only methods, all of the cross-domain methods achieve significant improvements, indicating the importance of source-domain information. Compared to methods with randomly initialized word embeddings, including BILSTM (Random) and MULTI-TASK (Random), BILSTM and MULTI-TASK (LSTM) that use pretrained word embeddings are able to obtain large performance enhancements. Combining with a pretrained LM is also helpful for the cross-domain task, since BERT-based methods perform best among all methods. 
\end{itemize}

\noindent%
In summary, the C-E relation encoder and bi-feedback scheme enhance the performance. Encoding relations between elements and clauses, as well as capturing interactions between clause classification and contract element extraction, are both beneficial for the cross-domain CEE problem. 

\negskip
\subsection{Cross-domain NER}
\label{subsection:NER}
\begin{table}[t]
  \small
  \caption{Results on three cross-domain NER datasets. Significant improvements against the best baseline with the same context encoder are marked with $\ast$ (t-test, $p < 0.05$).}
  \label{tab:cross-domain NER results}
  \begin{tabular}{lccc}
    \toprule
    \multirow{2}{*}{Model} & \multicolumn{3}{c}{Datasets (F1)} \\
    \cmidrule(r){2-4}
    ~ & BioNLP & Broad Twitter & Twitter \\
    \midrule
    BILSTM  & 79.24  & 72.98 & 77.18 \\
    MULTI-CELL LSTM  & 78.76 &72.54 & 77.05 \\
    \midrule
    MULTI-TASK (LSTM)  & 81.06 &73.84 & 79.55 \\
    MULTI-TASK+PGN  & 81.17 & 73.70 & 80.07 \\
    \citet{DBLP:conf/acl/JiaXZ19}  & 79.86 & -- & -- \\
    \citet{DBLP:conf/acl/JiaZ20}  & 83.12 & 74.82 & 81.37 \\
    \midrule
    BERT-BASE  & -- & 77.28 & 83.77 \\
    BIOBERT-BASE  & 85.72 & -- & --\\
    \citet{DBLP:conf/acl/JiaZ20} (BERT)  & 86.96 & 78.43 & 85.80 \\
    \midrule
    \ac{Bi-FLEET} (LSTM) & $82.32$\rlap{$^{\ast}$} & $74.91$\rlap{$^{\ast}$} & $80.78$\rlap{$^{\ast}$}\\
    \ac{Bi-FLEET} (PGN)  & $82.46$\rlap{$^{\ast}$} & $75.11$\rlap{$^{\ast}$} & $81.35$\rlap{$^{\ast}$}\\
    \ac{Bi-FLEET} (MULTI-CELL) & $84.76$\rlap{$^{\ast}$} & $76.02$\rlap{$^{\ast}$} & $82.24$\rlap{$^{\ast}$}\\
    \ac{Bi-FLEET} (BERT) & $\mathbf{89.02}$\rlap{$^{\ast}$} & $\mathbf{80.71}$\rlap{$^{\ast}$} & $\mathbf{87.73}$\rlap{$^{\ast}$}\\ 
  \bottomrule
\end{tabular}
\end{table}
To investigate the generality of \ac{Bi-FLEET}, we turn to RQ2 and report on experiments on the cross-domain NER datasets. F1-scores are adopted here to evaluate the overall performance. As baselines, we use the results reported in \citep{DBLP:conf/acl/JiaZ20}. Based on the results in Table~\ref{tab:cross-domain NER results}, we arrive at the following conclusions:

\begin{itemize}[leftmargin=*,nosep]
  \item Even if labels are generated in an unsupervised manner, Bi-FLEET can still effectively transfer named entity recognizers.
  \item Bi-FLEET outperforms state-of-the-art cross-domain NER methods. Lacking source-domain information, target-domain only methods can hardly handle the cross-domain NER task. Compared to cross-domain and BERT-based NER methods, methods based on the Bi-FLEET framework attain the highest F1-scores on all three datasets.
\end{itemize}

\noindent%
In summary, the Bi-FLEET framework cannot only effectively address the cross-domain \ac{CEE} problem, but also achieves state-of-the-art results on the cross-domain \ac{NER} task.

%% file: sections/analysis.tex

\section{Analysis}
\label{subsec:extensive analysis}
Now that we have answered our research questions, we take a closer the look at Bi-FLEET to analyze its performance. We examine how the \ac{C-E} relation encoder and bi-feedback scheme contribute to its performance, how the amount of target-domain data influences the performance, and how performance varies across element types.

\begin{table*}[t]
  \small
  \caption{Examples from the I2C dataset. Red and green represent incorrect and correct contract elements, respectively.}
  \label{tab:case studies}
  \begin{tabular}{l l}
    \toprule
    Models & Sentences \\
    \midrule
    \multirow{3}*{MULTI-TASK (LSTM)} & ``Party B shall return the remuneration received and pay \colorbox{red}{10\%} (\textbf{Payment Rate}) of the total remuneration \\  
    ~ & as liquidated damages.'' (\textbf{Indemnity Clause})\\
    ~ &``\colorbox{red}{According to the construction area} (\textbf{O}), the unit price of the commercial housing is XXX yuan per square meter''\\
    \midrule
    \multirow{4}*{Bi-FLEET (LSTM)} & ``Party B shall return the remuneration received and pay \colorbox{green}{10\%} (\textbf{Compensation Ratio}) of  the total remuneration \\
    ~ & as liquidated damages.''  (\textbf{Indemnity Clause})\\
    ~ & ``\colorbox{green}{According to the construction area} (\textbf{Calculation Standard}), the unit price of the commercial housing is XXX yuan\\
    ~ & per square meter''\\  
  \bottomrule
\end{tabular}
\end{table*}

\subsection{Ablation studies}
\label{subsection:ablation studies}
We conduct ablation studies on both cross-domain CEE and NER datasets. The results are shown in Table~\ref{tab:ablation studies}. When we only ablate the \ac{C-E} relation encoder (``- \ac{C-E} graph''), type representations $\mathbf{cla}$ and $\mathbf{ele}$ are randomly initialized. In that case, F1-scores over all of the three datasets and model variants suffer a severe drop. Since the \ac{C-E} relation encoder cannot be directly incorporated, we cannot remove the bi-feedback scheme alone. When we ablate both the \ac{C-E} relation encoder and bi-feedback scheme (``- Bi-feedback''), our models obtain a similar performance as baselines (MULTI-TASK (LSTM), MULTI-TASK+PGN and BERT). In short, both the \ac{C-E} relation encoder and the bi-feedback scheme contribute to the improvements in performance on the cross-domain CEE and NER tasks. 
  
\begin{table}[t]
  \small
  \setlength\tabcolsep{15pt}
  \caption{Ablation studies on the BioNLP, I2C and C2I datasets.}
  \label{tab:ablation studies}
  \begin{tabular}{lccc}
    \toprule
    \multirow{2}{*}{Model} & \multicolumn{3}{c}{Datasets (F1)} \\
    \cmidrule(r){2-4}
    ~ & BioNLP  & I2C & C2I\\
    \midrule
    \ac{Bi-FLEET} (LSTM)  & \textbf{82.32}  & \textbf{68.07} & \textbf{71.44}  \\
    \quad\quad- C-E graph  & 81.78  & 66.95 & 70.03   \\
    \quad\quad- Bi-feedback  & 81.03  & 65.29 & 68.59   \\
    \midrule
    \ac{Bi-FLEET} (PGN)  & \textbf{82.46}  & \textbf{68.43} & \textbf{71.69}  \\
    \quad\quad- C-E graph  & 81.85  & 67.37 & 70.15   \\
    \quad\quad- Bi-feedback  & 81.07  & 65.58 & 68.72   \\
    \midrule
    \ac{Bi-FLEET} (BERT)  & \textbf{89.02}  & \textbf{71.92} & \textbf{74.49}  \\
    \quad\quad- C-E graph  & 88.38  & 70.58 & 73.31   \\
    \quad\quad- Bi-feedback  & 86.89  & 69.32 & 72.06   \\
  \bottomrule
\end{tabular}
\end{table}

\subsection{Influence of target-domain data}
\label{subsection: influence of targe-domain data}
\begin{figure}
  \centering
  \includegraphics[width=0.75\linewidth]{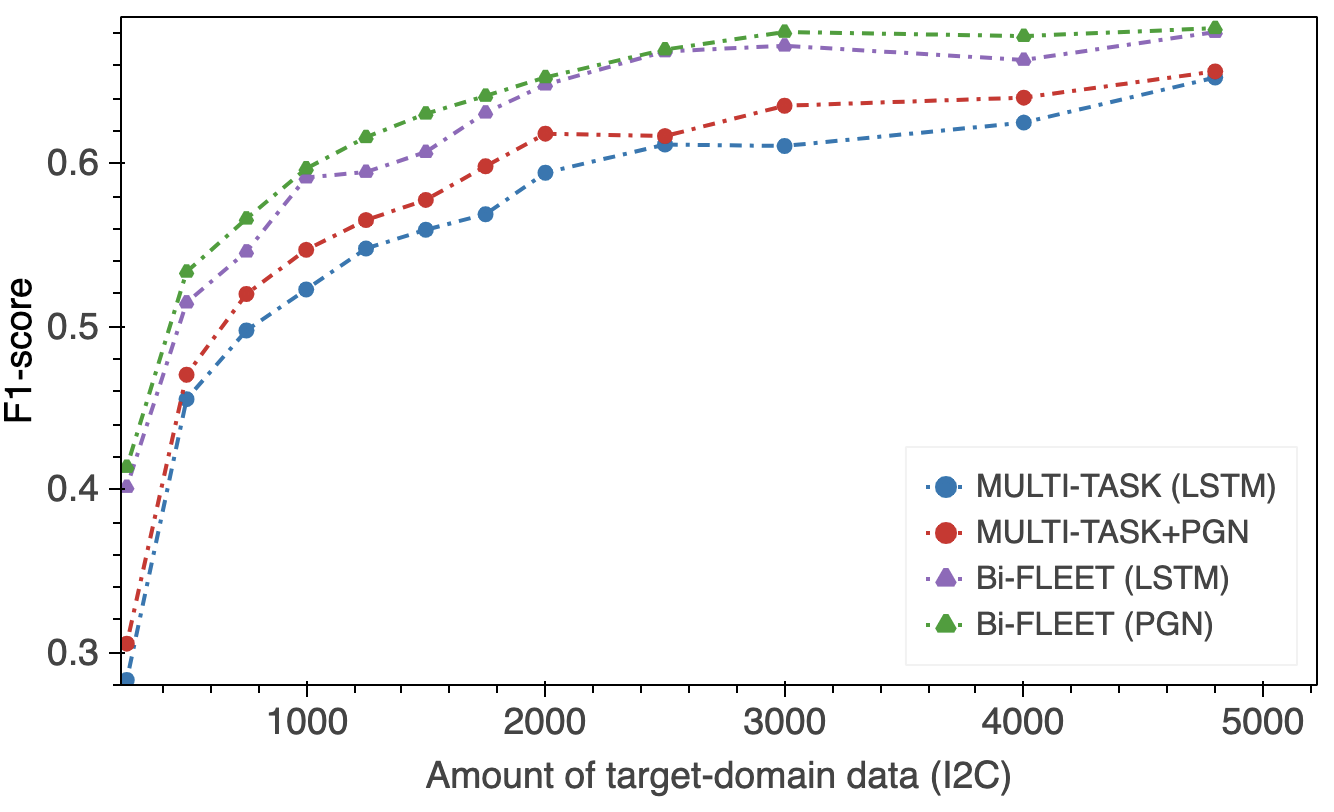}
  \caption{Influence of target-domain data.} 
  \label{fig:influence of target-domain data}
\end{figure}

Next, we study the influence of target-domain data on the I2C dataset. We compare the F1-scores of the baselines and Bi-FLEET with different amounts of target-domain data in 100 training epochs. 
We keep ``Neg'' sentences unchanged and adjust the number of ``Pos'' sentences, since ``Neg'' sentences do not contain any type of elements. 
See Fig.~\ref{fig:influence of target-domain data} for the results.  Initially, Bi-FLEET obtains a large improvement of more than 10\% over baselines. The gap between \ac{Bi-FLEET} and baselines becomes smaller with the increase in target-domain data. Both Bi-FLEET (LSTM) and Bi-FLEET (PGN) outperform the baselines with varying numbers of target-domain data, which demonstrates the effectiveness of our methods.  

\smallnegskip
\subsection{Fined-grained comparisons}
\label{subsection:fined-grained comparisons}
To understand the performance of Bi-FLEET at the element type level, we make fine-grained comparisons on the I2C dataset. As shown in Table~\ref{tab:fine-grained comparisons on I2C}, we analyze five element types: invoice type (IT), arbitration commission (AC), currency of payment (CP), name of subject matter (NSM), and payment period (PP). Models based on Bi-FLEET accomplish significant F1-score improvements over the baselines, demonstrating that modeling \ac{C-E} relations and uncovering the bi-feedback scheme are helpful for reducing confusion between element types.

\begin{table}[t]
  \small
  \caption{Fine-grained comparisons on the I2C dataset.}
  \label{tab:fine-grained comparisons on I2C}
  \begin{tabular}{l ccc cc}
    \toprule
    \multirow{2}{*}{Model} & \multicolumn{5}{c}{Element types (F1)} \\
    \cmidrule(r){2-6}
    ~ & IT & AC & CP & NSM & PP \\
    \midrule
    BILSTM   & 53.46  & 60.31 & 62.75  & 64.14  & 67.20  \\
    MULTI-TASK (LSTM)  & 57.92 & 63.61 & 63.80 & 65.43 & 67.39  \\
    MULTI-TASK+PGN  & 58.05 & 63.38 & 64.26 & 65.65 & 67.92 \\
    \midrule
    \ac{Bi-FLEET} (LSTM)  & 62.86 & 63.94 & 67.68 & 68.82 & 68.15  \\    
    \ac{Bi-FLEET} (PGN)  & \textbf{63.74} & \textbf{64.03} & \textbf{67.94} & \textbf{69.42} & \textbf{69.10}  \\
  \bottomrule
\end{tabular}
\end{table}

\subsection{Case studies}
\label{subsection:case studies}
Table~\ref{tab:case studies} provides examples from the I2C dataset. ``10\%'' is a compensation ratio, and ``According to the construction area'' is a calculation standard. Misled by the keywords ``pay'' and ``total remuneration'', MULTI-TASK (LSTM) misjudges ``10\%'' as a payment rate. Due to the data sparsity for the calculation standard, MULTI-TASK (LSTM) is not able to recognize ``According to the construction area''. By uncovering interactions between the \acf{CC} and \acf{CEE} tasks, Bi-FLEET can correctly extract the compensation rate in the indemnity clause. Identifying invariant knowledge about element types makes it much easier to transfer long-tail element types.  

%% file: sections/conclusion.tex

\section{conclusions}
\label{sec:conclusion}
In this paper, we have investigated the new problem of cross-domain \acl{CEE}; solutions to this problem can support clause retrieval and other legal search tasks. Compared to cross-domain \acl{NER}, there are two main challenges: more fine-grained element types and larger extraction zones. To address these challenges, we have proposed a framework named Bi-FLEET that captures invariant relations between clauses and elements, and we have designed a bidirectional feedback scheme between the \ac{CEE} and \acl{CC} tasks. Experimental results on cross-domain \ac{CEE} and \ac{NER} datasets show that Bi-FLEET significantly outperforms state-of-the-art baselines, with a high degree of generalizability.

A limitation of Bi-FLEET is that it can only extract elements sentence by sentence. In future work, we intend to consider the clause-level features, and explore relations between elements.

%% file: sections/acknowledgement.tex
\begin{acks}
We thank our reviewers for their feedback. This work was supported by the National Key R\&D Program of China with grant No. 2020YFB1406704, the Natural Science Foundation of China (61972234, 61902219, 62072279), 
the Key Scientific and Technological Innovation Program of Shandong Province (2019JZZY010129), 
Alibaba Group through Alibaba Research Intern Program, the Tencent WeChat Rhino-Bird Focused Research Program (JR-WXG-2021411), the Fundamental Research Funds of Shandong University, the Hybrid Intelligence Center, 
a 10-year program funded by the Dutch Ministry of Education, Culture and Science through 
the Netherlands Organisation for Scientific Research, \url{https://hybrid-intelligence-centre.nl}. All content represents the opinion of the authors, which is not necessarily shared or endorsed by their respective employers and/or sponsors. 
\end{acks}

%% file: main.bbl

\begin{thebibliography}{44}


\ifx \showCODEN    \undefined \def \showCODEN     #1{\unskip}     \fi
\ifx \showDOI      \undefined \def \showDOI       #1{#1}\fi
\ifx \showISBNx    \undefined \def \showISBNx     #1{\unskip}     \fi
\ifx \showISBNxiii \undefined \def \showISBNxiii  #1{\unskip}     \fi
\ifx \showISSN     \undefined \def \showISSN      #1{\unskip}     \fi
\ifx \showLCCN     \undefined \def \showLCCN      #1{\unskip}     \fi
\ifx \shownote     \undefined \def \shownote      #1{#1}          \fi
\ifx \showarticletitle \undefined \def \showarticletitle #1{#1}   \fi
\ifx \showURL      \undefined \def \showURL       {\relax}        \fi
\providecommand\bibfield[2]{#2}
\providecommand\bibinfo[2]{#2}
\providecommand\natexlab[1]{#1}
\providecommand\showeprint[2][]{arXiv:#2}

\bibitem[\protect\citeauthoryear{Azzopardi, Gatt, and Pace}{Azzopardi
  et~al\mbox{.}}{2016}]%
        {azzopardi2016integrating}
\bibfield{author}{\bibinfo{person}{Shaun Azzopardi}, \bibinfo{person}{Albert
  Gatt}, {and} \bibinfo{person}{Gordon~J Pace}.}
  \bibinfo{year}{2016}\natexlab{}.
\newblock \showarticletitle{Integrating natural language and formal analysis
  for legal documents}. In \bibinfo{booktitle}{\emph{10th Conference on
  Language Technologies and Digital Humanities}}. \bibinfo{pages}{1--4}.
\newblock


\bibitem[\protect\citeauthoryear{Borchmann, Wisniewski, Gretkowski, Kosmala,
  Jurkiewicz, Szalkiewicz, Palka, Kaczmarek, Kaliska, and Gralinski}{Borchmann
  et~al\mbox{.}}{2020}]%
        {DBLP:conf/emnlp/BorchmannWGKJSP20}
\bibfield{author}{\bibinfo{person}{Lukasz Borchmann}, \bibinfo{person}{Dawid
  Wisniewski}, \bibinfo{person}{Andrzej Gretkowski}, \bibinfo{person}{Izabela
  Kosmala}, \bibinfo{person}{Dawid Jurkiewicz}, \bibinfo{person}{Lukasz
  Szalkiewicz}, \bibinfo{person}{Gabriela Palka}, \bibinfo{person}{Karol
  Kaczmarek}, \bibinfo{person}{Agnieszka Kaliska}, {and} \bibinfo{person}{Filip
  Gralinski}.} \bibinfo{year}{2020}\natexlab{}.
\newblock \showarticletitle{Contract discovery: Dataset and a few-shot semantic
  retrieval challenge with competitive baselines}. In
  \bibinfo{booktitle}{\emph{Proceedings of the 2020 Conference on Empirical
  Methods in Natural Language Processing: Findings}}.
  \bibinfo{pages}{4254--4268}.
\newblock


\bibitem[\protect\citeauthoryear{Cao, Chen, Liu, Zhao, and Liu}{Cao
  et~al\mbox{.}}{2018}]%
        {DBLP:conf/emnlp/Cao000L18}
\bibfield{author}{\bibinfo{person}{Pengfei Cao}, \bibinfo{person}{Yubo Chen},
  \bibinfo{person}{Kang Liu}, \bibinfo{person}{Jun Zhao}, {and}
  \bibinfo{person}{Shengping Liu}.} \bibinfo{year}{2018}\natexlab{}.
\newblock \showarticletitle{Adversarial transfer learning for Chinese named
  entity recognition with self-attention mechanism}. In
  \bibinfo{booktitle}{\emph{Proceedings of the 2018 Conference on Empirical
  Methods in Natural Language Processing}}. \bibinfo{publisher}{Association for
  Computational Linguistics}, \bibinfo{pages}{182--192}.
\newblock


\bibitem[\protect\citeauthoryear{Chalkidis and Androutsopoulos}{Chalkidis and
  Androutsopoulos}{2017}]%
        {chalkidis2017deep}
\bibfield{author}{\bibinfo{person}{Ilias Chalkidis} {and} \bibinfo{person}{Ion
  Androutsopoulos}.} \bibinfo{year}{2017}\natexlab{}.
\newblock \showarticletitle{A deep learning approach to contract element
  extraction}. In \bibinfo{booktitle}{\emph{Legal Knowledge and Information
  Systems - {JURIX} 2017: The Thirtieth Annual Conference}}.
  \bibinfo{pages}{155--164}.
\newblock


\bibitem[\protect\citeauthoryear{Chalkidis, Androutsopoulos, and
  Michos}{Chalkidis et~al\mbox{.}}{2017}]%
        {chalkidis2017extracting}
\bibfield{author}{\bibinfo{person}{Ilias Chalkidis}, \bibinfo{person}{Ion
  Androutsopoulos}, {and} \bibinfo{person}{Achilleas Michos}.}
  \bibinfo{year}{2017}\natexlab{}.
\newblock \showarticletitle{Extracting contract elements}. In
  \bibinfo{booktitle}{\emph{Proceedings of the 16th edition of the
  International Conference on Articial Intelligence and Law}}.
  \bibinfo{pages}{19--28}.
\newblock


\bibitem[\protect\citeauthoryear{Chalkidis, Fergadiotis, Malakasiotis, and
  Androutsopoulos}{Chalkidis et~al\mbox{.}}{2019}]%
        {chalkidis2019neural}
\bibfield{author}{\bibinfo{person}{Ilias Chalkidis}, \bibinfo{person}{Manos
  Fergadiotis}, \bibinfo{person}{Prodromos Malakasiotis}, {and}
  \bibinfo{person}{Ion Androutsopoulos}.} \bibinfo{year}{2019}\natexlab{}.
\newblock \showarticletitle{Neural contract element extraction revisited}. In
  \bibinfo{booktitle}{\emph{Workshop on Document Intelligence at NeurIPS
  2019}}.
\newblock


\bibitem[\protect\citeauthoryear{Chiu, Crichton, Korhonen, and Pyysalo}{Chiu
  et~al\mbox{.}}{2016}]%
        {DBLP:conf/bionlp/ChiuCKP16}
\bibfield{author}{\bibinfo{person}{Billy Chiu}, \bibinfo{person}{Gamal K.~O.
  Crichton}, \bibinfo{person}{Anna Korhonen}, {and} \bibinfo{person}{Sampo
  Pyysalo}.} \bibinfo{year}{2016}\natexlab{}.
\newblock \showarticletitle{How to train good word embeddings for biomedical
  {NLP}}. In \bibinfo{booktitle}{\emph{Proceedings of the 15th Workshop on
  Biomedical Natural Language Processing}}. \bibinfo{publisher}{Association for
  Computational Linguistics}, \bibinfo{pages}{166--174}.
\newblock


\bibitem[\protect\citeauthoryear{Chu}{Chu}{2011}]%
        {DBLP:journals/jd/Chu11}
\bibfield{author}{\bibinfo{person}{Heting Chu}.}
  \bibinfo{year}{2011}\natexlab{}.
\newblock \showarticletitle{Factors affecting relevance judgment: A report from
  {TREC} Legal track}.
\newblock \bibinfo{journal}{\emph{J. Documentation}} \bibinfo{volume}{67},
  \bibinfo{number}{2} (\bibinfo{year}{2011}), \bibinfo{pages}{264--278}.
\newblock


\bibitem[\protect\citeauthoryear{Curtotti and Mccreath}{Curtotti and
  Mccreath}{2010}]%
        {curtotti2010corpus}
\bibfield{author}{\bibinfo{person}{Michael Curtotti} {and}
  \bibinfo{person}{Eric Mccreath}.} \bibinfo{year}{2010}\natexlab{}.
\newblock \showarticletitle{Corpus based classification of text in Australian
  contracts}. In \bibinfo{booktitle}{\emph{Proceedings of the Australasian
  Language Technology Association Workshop}}.
\newblock


\bibitem[\protect\citeauthoryear{Derczynski, Bontcheva, and Roberts}{Derczynski
  et~al\mbox{.}}{2016}]%
        {DBLP:conf/coling/DerczynskiBR16}
\bibfield{author}{\bibinfo{person}{Leon Derczynski}, \bibinfo{person}{Kalina
  Bontcheva}, {and} \bibinfo{person}{Ian Roberts}.}
  \bibinfo{year}{2016}\natexlab{}.
\newblock \showarticletitle{Broad Twitter corpus: {A} diverse named entity
  recognition resource}. In \bibinfo{booktitle}{\emph{{COLING} 2016, 26th
  International Conference on Computational Linguistics}}.
  \bibinfo{pages}{1169--1179}.
\newblock


\bibitem[\protect\citeauthoryear{Devlin, Chang, Lee, and Toutanova}{Devlin
  et~al\mbox{.}}{2019}]%
        {devlin2018bert}
\bibfield{author}{\bibinfo{person}{Jacob Devlin}, \bibinfo{person}{Ming-Wei
  Chang}, \bibinfo{person}{Kenton Lee}, {and} \bibinfo{person}{Kristina
  Toutanova}.} \bibinfo{year}{2019}\natexlab{}.
\newblock \showarticletitle{Bert: Pre-training of deep bidirectional
  transformers for language understanding}. In
  \bibinfo{booktitle}{\emph{Proceedings of the 2019 Conference of the North
  American Chapter of the Association for Computational Linguistics: Human
  Language Technologies}}. \bibinfo{pages}{4171--4186}.
\newblock


\bibitem[\protect\citeauthoryear{Do, Nguyen, Tran, Nguyen, and Nguyen}{Do
  et~al\mbox{.}}{2017}]%
        {do2017legal}
\bibfield{author}{\bibinfo{person}{Phong-Khac Do}, \bibinfo{person}{Huy-Tien
  Nguyen}, \bibinfo{person}{Chien-Xuan Tran}, \bibinfo{person}{Minh-Tien
  Nguyen}, {and} \bibinfo{person}{Minh-Le Nguyen}.}
  \bibinfo{year}{2017}\natexlab{}.
\newblock \showarticletitle{Legal question answering using ranking SVM and deep
  convolutional neural network}.
\newblock \bibinfo{journal}{\emph{arXiv preprint arXiv:1703.05320}}
  (\bibinfo{year}{2017}).
\newblock


\bibitem[\protect\citeauthoryear{Feng, He, Tang, and Chua}{Feng
  et~al\mbox{.}}{2019}]%
        {feng2019graph}
\bibfield{author}{\bibinfo{person}{Fuli Feng}, \bibinfo{person}{Xiangnan He},
  \bibinfo{person}{Jie Tang}, {and} \bibinfo{person}{Tat-Seng Chua}.}
  \bibinfo{year}{2019}\natexlab{}.
\newblock \showarticletitle{Graph adversarial training: Dynamically
  regularizing based on graph structure}.
\newblock \bibinfo{journal}{\emph{IEEE Transactions on Knowledge and Data
  Engineering}} (\bibinfo{year}{2019}).
\newblock


\bibitem[\protect\citeauthoryear{Feng, Huang, He, Xin, Wang, and Chua}{Feng
  et~al\mbox{.}}{2021}]%
        {feng2021causalgcn}
\bibfield{author}{\bibinfo{person}{Fuli Feng}, \bibinfo{person}{Weiran Huang},
  \bibinfo{person}{Xiangnan He}, \bibinfo{person}{Xin Xin},
  \bibinfo{person}{Qifan Wang}, {and} \bibinfo{person}{Tat-Seng Chua}.}
  \bibinfo{year}{2021}\natexlab{}.
\newblock \showarticletitle{Should Graph Convolution Trust Neighbors? A Simple
  Causal Inference Method}. In \bibinfo{booktitle}{\emph{Proceedings of the
  44th International ACM SIGIR Conference on Research and Development in
  Information Retrieval}}.
\newblock


\bibitem[\protect\citeauthoryear{Garc{\'\i}a-Constantino, Atkinson, Bollegala,
  Chapman, Coenen, Roberts, and Robson}{Garc{\'\i}a-Constantino
  et~al\mbox{.}}{2017}]%
        {garcia2017cliel}
\bibfield{author}{\bibinfo{person}{Mat{\'\i}as Garc{\'\i}a-Constantino},
  \bibinfo{person}{Katie Atkinson}, \bibinfo{person}{Danushka Bollegala},
  \bibinfo{person}{Karl Chapman}, \bibinfo{person}{Frans Coenen},
  \bibinfo{person}{Claire Roberts}, {and} \bibinfo{person}{Katy Robson}.}
  \bibinfo{year}{2017}\natexlab{}.
\newblock \showarticletitle{CLIEL: Context-based information extraction from
  commercial law documents}. In \bibinfo{booktitle}{\emph{Proceedings of the
  16th edition of the International Conference on Articial Intelligence and
  Law}}. \bibinfo{pages}{79--87}.
\newblock


\bibitem[\protect\citeauthoryear{Gimpel and Smith}{Gimpel and Smith}{2010}]%
        {DBLP:conf/naacl/GimpelS10}
\bibfield{author}{\bibinfo{person}{Kevin Gimpel} {and} \bibinfo{person}{Noah~A.
  Smith}.} \bibinfo{year}{2010}\natexlab{}.
\newblock \showarticletitle{Softmax-margin CRFs: Training log-linear models
  with cost functions}. In \bibinfo{booktitle}{\emph{Human Language
  Technologies: Conference of the North American Chapter of the Association of
  Computational Linguistics}}. \bibinfo{pages}{733--736}.
\newblock


\bibitem[\protect\citeauthoryear{Graves and Schmidhuber}{Graves and
  Schmidhuber}{2005}]%
        {graves2005framewise}
\bibfield{author}{\bibinfo{person}{Alex Graves} {and}
  \bibinfo{person}{J{\"u}rgen Schmidhuber}.} \bibinfo{year}{2005}\natexlab{}.
\newblock \showarticletitle{Framewise phoneme classification with bidirectional
  LSTM and other neural network architectures}.
\newblock \bibinfo{journal}{\emph{Neural networks}} \bibinfo{volume}{18},
  \bibinfo{number}{5-6} (\bibinfo{year}{2005}), \bibinfo{pages}{602--610}.
\newblock


\bibitem[\protect\citeauthoryear{Indukuri and Krishna}{Indukuri and
  Krishna}{2010}]%
        {indukuri2010mining}
\bibfield{author}{\bibinfo{person}{Kishore~Varma Indukuri} {and}
  \bibinfo{person}{P~Radha Krishna}.} \bibinfo{year}{2010}\natexlab{}.
\newblock \showarticletitle{Mining e-contract documents to classify clauses}.
  In \bibinfo{booktitle}{\emph{Proceedings of the 3rd Bangalore Annual Compute
  Conference}}. \bibinfo{pages}{1--5}.
\newblock


\bibitem[\protect\citeauthoryear{Jia, Xiao, and Zhang}{Jia
  et~al\mbox{.}}{2019}]%
        {DBLP:conf/acl/JiaXZ19}
\bibfield{author}{\bibinfo{person}{Chen Jia}, \bibinfo{person}{Liang Xiao},
  {and} \bibinfo{person}{Yue Zhang}.} \bibinfo{year}{2019}\natexlab{}.
\newblock \showarticletitle{Cross-domain {NER} using cross-domain language
  modeling}. In \bibinfo{booktitle}{\emph{Proceedings of the 57th Conference of
  the Association for Computational Linguistics}}. \bibinfo{pages}{2464--2474}.
\newblock


\bibitem[\protect\citeauthoryear{Jia and Zhang}{Jia and Zhang}{2020}]%
        {DBLP:conf/acl/JiaZ20}
\bibfield{author}{\bibinfo{person}{Chen Jia} {and} \bibinfo{person}{Yue
  Zhang}.} \bibinfo{year}{2020}\natexlab{}.
\newblock \showarticletitle{Multi-Cell compositional {LSTM} for {NER} domain
  adaptation}. In \bibinfo{booktitle}{\emph{Proceedings of the 58th Annual
  Meeting of the Association for Computational Linguistics}}.
  \bibinfo{pages}{5906--5917}.
\newblock


\bibitem[\protect\citeauthoryear{Kano, Kim, Goebel, and Satoh}{Kano
  et~al\mbox{.}}{2017}]%
        {DBLP:conf/icail/KanoKGS17}
\bibfield{author}{\bibinfo{person}{Yoshinobu Kano}, \bibinfo{person}{Mi{-}Young
  Kim}, \bibinfo{person}{Randy Goebel}, {and} \bibinfo{person}{Ken Satoh}.}
  \bibinfo{year}{2017}\natexlab{}.
\newblock \showarticletitle{Overview of {COLIEE} 2017}. In
  \bibinfo{booktitle}{\emph{4th Competition on Legal Information Extraction and
  Entailment, held in conjunction with the 16th International Conference on
  Artificial Intelligence and Law}}, Vol.~\bibinfo{volume}{47}.
  \bibinfo{pages}{1--8}.
\newblock


\bibitem[\protect\citeauthoryear{Kien, Nguyen, Bach, Tran, Nguyen, and
  Phuong}{Kien et~al\mbox{.}}{2020}]%
        {DBLP:conf/coling/KienNBTNP20}
\bibfield{author}{\bibinfo{person}{Phi~Manh Kien}, \bibinfo{person}{Ha{-}Thanh
  Nguyen}, \bibinfo{person}{Ngo~Xuan Bach}, \bibinfo{person}{Vu Tran},
  \bibinfo{person}{Minh~Le Nguyen}, {and} \bibinfo{person}{Tu~Minh Phuong}.}
  \bibinfo{year}{2020}\natexlab{}.
\newblock \showarticletitle{Answering legal questions by learning neural
  attentive rext representation}. In \bibinfo{booktitle}{\emph{Proceedings of
  the 28th International Conference on Computational Linguistics}}.
  \bibinfo{pages}{988--998}.
\newblock


\bibitem[\protect\citeauthoryear{Lee, Yoon, Kim, Kim, Kim, So, and Kang}{Lee
  et~al\mbox{.}}{2020}]%
        {DBLP:journals/bioinformatics/LeeYKKKSK20}
\bibfield{author}{\bibinfo{person}{Jinhyuk Lee}, \bibinfo{person}{Wonjin Yoon},
  \bibinfo{person}{Sungdong Kim}, \bibinfo{person}{Donghyeon Kim},
  \bibinfo{person}{Sunkyu Kim}, \bibinfo{person}{Chan~Ho So}, {and}
  \bibinfo{person}{Jaewoo Kang}.} \bibinfo{year}{2020}\natexlab{}.
\newblock \showarticletitle{BioBERT: A pre-trained biomedical language
  representation model for biomedical text mining}.
\newblock \bibinfo{journal}{\emph{Bioinform.}} \bibinfo{volume}{36},
  \bibinfo{number}{4} (\bibinfo{year}{2020}), \bibinfo{pages}{1234--1240}.
\newblock


\bibitem[\protect\citeauthoryear{Lee, Dernoncourt, and Szolovits}{Lee
  et~al\mbox{.}}{2018}]%
        {DBLP:conf/lrec/LeeDS18}
\bibfield{author}{\bibinfo{person}{Ji~Young Lee}, \bibinfo{person}{Franck
  Dernoncourt}, {and} \bibinfo{person}{Peter Szolovits}.}
  \bibinfo{year}{2018}\natexlab{}.
\newblock \showarticletitle{Transfer learning for named-entity recognition with
  neural networks}. In \bibinfo{booktitle}{\emph{Proceedings of the 11th
  International Conference on Language Resources and Evaluation}}.
\newblock


\bibitem[\protect\citeauthoryear{Li, Sun, Han, and Li}{Li
  et~al\mbox{.}}{2020}]%
        {li2020survey}
\bibfield{author}{\bibinfo{person}{Jing Li}, \bibinfo{person}{Aixin Sun},
  \bibinfo{person}{Jianglei Han}, {and} \bibinfo{person}{Chenliang Li}.}
  \bibinfo{year}{2020}\natexlab{}.
\newblock \showarticletitle{A survey on deep learning for named entity
  recognition}.
\newblock \bibinfo{journal}{\emph{IEEE Transactions on Knowledge and Data
  Engineering}} (\bibinfo{year}{2020}).
\newblock


\bibitem[\protect\citeauthoryear{Lin and Lu}{Lin and Lu}{2018}]%
        {DBLP:conf/emnlp/LinL18}
\bibfield{author}{\bibinfo{person}{Bill~Yuchen Lin} {and} \bibinfo{person}{Wei
  Lu}.} \bibinfo{year}{2018}\natexlab{}.
\newblock \showarticletitle{Neural adaptation layers for cross-domain named
  entity recognition}. In \bibinfo{booktitle}{\emph{Proceedings of the 2018
  Conference on Empirical Methods in Natural Language Processing}}.
  \bibinfo{pages}{2012--2022}.
\newblock


\bibitem[\protect\citeauthoryear{Lu, Neves, Carvalho, Zhang, and Ji}{Lu
  et~al\mbox{.}}{2018}]%
        {DBLP:conf/acl/JiZCLN18}
\bibfield{author}{\bibinfo{person}{Di Lu}, \bibinfo{person}{Leonardo Neves},
  \bibinfo{person}{Vitor Carvalho}, \bibinfo{person}{Ning Zhang}, {and}
  \bibinfo{person}{Heng Ji}.} \bibinfo{year}{2018}\natexlab{}.
\newblock \showarticletitle{Visual attention model for name tagging in
  multimodal social media}. In \bibinfo{booktitle}{\emph{Proceedings of the
  56th Annual Meeting of the Association for Computational Linguistics}}.
  \bibinfo{pages}{1990--1999}.
\newblock


\bibitem[\protect\citeauthoryear{Ma and Hovy}{Ma and Hovy}{2016}]%
        {DBLP:conf/acl/MaH16}
\bibfield{author}{\bibinfo{person}{Xuezhe Ma} {and} \bibinfo{person}{Eduard~H.
  Hovy}.} \bibinfo{year}{2016}\natexlab{}.
\newblock \showarticletitle{End-to-end sequence labeling via bi-directional
  LSTM-CNNs-CRF}. In \bibinfo{booktitle}{\emph{Proceedings of the 54th Annual
  Meeting of the Association for Computational Linguistics}}.
\newblock


\bibitem[\protect\citeauthoryear{Maxwell and Schafer}{Maxwell and
  Schafer}{2008}]%
        {DBLP:conf/jurix/MaxwellS08}
\bibfield{author}{\bibinfo{person}{K.~Tamsin Maxwell} {and}
  \bibinfo{person}{Burkhard Schafer}.} \bibinfo{year}{2008}\natexlab{}.
\newblock \showarticletitle{Concept and context in legal information
  retrieval}. In \bibinfo{booktitle}{\emph{{JURIX} 2008: The Twenty-First
  Annual Conference on Legal Knowledge and Information Systems}},
  \bibfield{editor}{\bibinfo{person}{Enrico Francesconi},
  \bibinfo{person}{Giovanni Sartor}, {and} \bibinfo{person}{Daniela Tiscornia}}
  (Eds.), Vol.~\bibinfo{volume}{189}. \bibinfo{pages}{63--72}.
\newblock


\bibitem[\protect\citeauthoryear{Milosevic, Gibson, Linington, Cole, and
  Kulkarni}{Milosevic et~al\mbox{.}}{2004}]%
        {milosevic2004design}
\bibfield{author}{\bibinfo{person}{Zoran Milosevic}, \bibinfo{person}{Simon
  Gibson}, \bibinfo{person}{Peter~F Linington}, \bibinfo{person}{James Cole},
  {and} \bibinfo{person}{Sachin Kulkarni}.} \bibinfo{year}{2004}\natexlab{}.
\newblock \showarticletitle{On design and implementation of a contract
  monitoring facility}. In \bibinfo{booktitle}{\emph{Proceedings of the 1st
  IEEE International Workshop on Electronic Contracting}}.
  \bibinfo{pages}{62--70}.
\newblock


\bibitem[\protect\citeauthoryear{N{\'{e}}dellec, Bossy, Kim, Kim, Ohta,
  Pyysalo, and Zweigenbaum}{N{\'{e}}dellec et~al\mbox{.}}{2013}]%
        {DBLP:conf/bionlp/NedellecBKKOPZ13}
\bibfield{author}{\bibinfo{person}{Claire N{\'{e}}dellec},
  \bibinfo{person}{Robert Bossy}, \bibinfo{person}{Jin{-}Dong Kim},
  \bibinfo{person}{Jung{-}Jae Kim}, \bibinfo{person}{Tomoko Ohta},
  \bibinfo{person}{Sampo Pyysalo}, {and} \bibinfo{person}{Pierre Zweigenbaum}.}
  \bibinfo{year}{2013}\natexlab{}.
\newblock \showarticletitle{Overview of BioNLP Shared Task 2013}. In
  \bibinfo{booktitle}{\emph{Proceedings of the BioNLP Shared Task 2013
  Workshop}}. \bibinfo{publisher}{Association for Computational Linguistics},
  \bibinfo{pages}{1--7}.
\newblock


\bibitem[\protect\citeauthoryear{Pan, Toh, and Su}{Pan et~al\mbox{.}}{2013}]%
        {pan2013transfer}
\bibfield{author}{\bibinfo{person}{Sinno~Jialin Pan}, \bibinfo{person}{Zhiqiang
  Toh}, {and} \bibinfo{person}{Jian Su}.} \bibinfo{year}{2013}\natexlab{}.
\newblock \showarticletitle{Transfer joint embedding for cross-domain named
  entity recognition}.
\newblock \bibinfo{journal}{\emph{ACM Transactions on Information Systems}}
  \bibinfo{volume}{31}, \bibinfo{number}{2} (\bibinfo{year}{2013}),
  \bibinfo{pages}{1--27}.
\newblock


\bibitem[\protect\citeauthoryear{Pennington, Socher, and Manning}{Pennington
  et~al\mbox{.}}{2014}]%
        {DBLP:conf/emnlp/PenningtonSM14}
\bibfield{author}{\bibinfo{person}{Jeffrey Pennington},
  \bibinfo{person}{Richard Socher}, {and} \bibinfo{person}{Christopher~D.
  Manning}.} \bibinfo{year}{2014}\natexlab{}.
\newblock \showarticletitle{Glove: Global vectors for word representation}. In
  \bibinfo{booktitle}{\emph{Proceedings of the 2014 Conference on Empirical
  Methods in Natural Language Processing}}. \bibinfo{pages}{1532--1543}.
\newblock


\bibitem[\protect\citeauthoryear{Perera, Jayawardana, Lakmal, and
  Perera}{Perera et~al\mbox{.}}{2018}]%
        {perera2018legal}
\bibfield{author}{\bibinfo{person}{Amal~Shehan Perera},
  \bibinfo{person}{Vindula Jayawardana}, \bibinfo{person}{Dimuthu Lakmal},
  {and} \bibinfo{person}{Madhavi Perera}.} \bibinfo{year}{2018}\natexlab{}.
\newblock \showarticletitle{Legal document retrieval using document vector
  embeddings and deep learning}. In \bibinfo{booktitle}{\emph{Intelligent
  Computing: Proceedings of the 2018 Computing Conference}},
  Vol.~\bibinfo{volume}{857}. \bibinfo{pages}{160}.
\newblock


\bibitem[\protect\citeauthoryear{Qu, Ferraro, Zhou, Hou, and Baldwin}{Qu
  et~al\mbox{.}}{2016}]%
        {DBLP:conf/emnlp/QuFZHB16}
\bibfield{author}{\bibinfo{person}{Lizhen Qu}, \bibinfo{person}{Gabriela
  Ferraro}, \bibinfo{person}{Liyuan Zhou}, \bibinfo{person}{Weiwei Hou}, {and}
  \bibinfo{person}{Timothy Baldwin}.} \bibinfo{year}{2016}\natexlab{}.
\newblock \showarticletitle{Named entity recognition for novel types by
  transfer learning}. In \bibinfo{booktitle}{\emph{Proceedings of the 2016
  Conference on Empirical Methods in Natural Language Processing}}.
  \bibinfo{pages}{899--905}.
\newblock


\bibitem[\protect\citeauthoryear{Sachan, Xie, Sachan, and Xing}{Sachan
  et~al\mbox{.}}{2018}]%
        {sachan2018effective}
\bibfield{author}{\bibinfo{person}{Devendra~Singh Sachan},
  \bibinfo{person}{Pengtao Xie}, \bibinfo{person}{Mrinmaya Sachan}, {and}
  \bibinfo{person}{Eric~P Xing}.} \bibinfo{year}{2018}\natexlab{}.
\newblock \showarticletitle{Effective use of bidirectional language modeling
  for transfer learning in biomedical named entity recognition}. In
  \bibinfo{booktitle}{\emph{Machine Learning for Healthcare Conference}}. PMLR,
  \bibinfo{pages}{383--402}.
\newblock


\bibitem[\protect\citeauthoryear{Sang and Meulder}{Sang and Meulder}{2003}]%
        {DBLP:conf/conll/SangM03}
\bibfield{author}{\bibinfo{person}{Erik F. Tjong~Kim Sang} {and}
  \bibinfo{person}{Fien~De Meulder}.} \bibinfo{year}{2003}\natexlab{}.
\newblock \showarticletitle{Introduction to the CoNLL-2003 Shared Task:
  Language-Independent Named Entity Recognition}. In
  \bibinfo{booktitle}{\emph{Proceedings of the 7th Conference on Natural
  Language Learning}}. \bibinfo{pages}{142--147}.
\newblock


\bibitem[\protect\citeauthoryear{Sun, Zhang, Ji, and Yang}{Sun
  et~al\mbox{.}}{2019}]%
        {sun2019toi}
\bibfield{author}{\bibinfo{person}{Lin Sun}, \bibinfo{person}{Kai Zhang},
  \bibinfo{person}{Fule Ji}, {and} \bibinfo{person}{Zhenhua Yang}.}
  \bibinfo{year}{2019}\natexlab{}.
\newblock \showarticletitle{Toi-CNN: A solution of information extraction on
  Chinese insurance policy}. In \bibinfo{booktitle}{\emph{Proceedings of the
  2019 Conference of the North American Chapter of the Association for
  Computational Linguistics: Human Language Technologies, Volume 2 (Industry
  Papers)}}. \bibinfo{pages}{174--181}.
\newblock


\bibitem[\protect\citeauthoryear{Tran, Nguyen, and Satoh}{Tran
  et~al\mbox{.}}{2019}]%
        {DBLP:conf/icail/TranNS19}
\bibfield{author}{\bibinfo{person}{Vu~D. Tran}, \bibinfo{person}{Minh~Le
  Nguyen}, {and} \bibinfo{person}{Ken Satoh}.} \bibinfo{year}{2019}\natexlab{}.
\newblock \showarticletitle{Building Legal Case Retrieval Systems with Lexical
  Matching and Summarization using {A} Pre-Trained Phrase Scoring Model}. In
  \bibinfo{booktitle}{\emph{Proceedings of the Seventeenth International
  Conference on Artificial Intelligence and Law}}. \bibinfo{pages}{275--282}.
\newblock


\bibitem[\protect\citeauthoryear{Tran, Nguyen, Tojo, and Satoh}{Tran
  et~al\mbox{.}}{2020}]%
        {DBLP:journals/ail/TranNTS20}
\bibfield{author}{\bibinfo{person}{Vu~D. Tran}, \bibinfo{person}{Minh~Le
  Nguyen}, \bibinfo{person}{Satoshi Tojo}, {and} \bibinfo{person}{Ken Satoh}.}
  \bibinfo{year}{2020}\natexlab{}.
\newblock \showarticletitle{Encoded summarization: summarizing documents into
  continuous vector space for legal case retrieval}.
\newblock \bibinfo{journal}{\emph{Artif. Intell. Law}} \bibinfo{volume}{28},
  \bibinfo{number}{4} (\bibinfo{year}{2020}), \bibinfo{pages}{441--467}.
\newblock


\bibitem[\protect\citeauthoryear{Wang, Ren, He, Zhang, and Hu}{Wang
  et~al\mbox{.}}{2019}]%
        {wang2019robust}
\bibfield{author}{\bibinfo{person}{Zihan Wang}, \bibinfo{person}{Zhaochun Ren},
  \bibinfo{person}{Chunyu He}, \bibinfo{person}{Peng Zhang}, {and}
  \bibinfo{person}{Yue Hu}.} \bibinfo{year}{2019}\natexlab{}.
\newblock \showarticletitle{Robust embedding with multi-level structures for
  link prediction}. In \bibinfo{booktitle}{\emph{Proceedings of the 28th
  International Joint Conference on Artificial Intelligence}}.
  \bibinfo{pages}{5240--5246}.
\newblock


\bibitem[\protect\citeauthoryear{Xu, Hu, Leskovec, and Jegelka}{Xu
  et~al\mbox{.}}{2019}]%
        {DBLP:conf/iclr/XuHLJ19}
\bibfield{author}{\bibinfo{person}{Keyulu Xu}, \bibinfo{person}{Weihua Hu},
  \bibinfo{person}{Jure Leskovec}, {and} \bibinfo{person}{Stefanie Jegelka}.}
  \bibinfo{year}{2019}\natexlab{}.
\newblock \showarticletitle{How powerful are graph neural networks?}. In
  \bibinfo{booktitle}{\emph{7th International Conference on Learning
  Representations}}.
\newblock


\bibitem[\protect\citeauthoryear{Yang and Zhang}{Yang and Zhang}{2018}]%
        {DBLP:conf/acl/YangZ18}
\bibfield{author}{\bibinfo{person}{Jie Yang} {and} \bibinfo{person}{Yue
  Zhang}.} \bibinfo{year}{2018}\natexlab{}.
\newblock \showarticletitle{{NCRF++:} An open-source neural sequence labeling
  toolkit}. In \bibinfo{booktitle}{\emph{Proceedings of the 56th Conference of
  the Association for Computational Linguistics}}. \bibinfo{pages}{74--79}.
\newblock


\bibitem[\protect\citeauthoryear{Yang, Salakhutdinov, and Cohen}{Yang
  et~al\mbox{.}}{2017}]%
        {DBLP:conf/iclr/YangSC17}
\bibfield{author}{\bibinfo{person}{Zhilin Yang}, \bibinfo{person}{Ruslan
  Salakhutdinov}, {and} \bibinfo{person}{William~W. Cohen}.}
  \bibinfo{year}{2017}\natexlab{}.
\newblock \showarticletitle{Transfer learning for sequence tagging with
  hierarchical recurrent networks}. In \bibinfo{booktitle}{\emph{5th
  International Conference on Learning Representations}}.
\newblock


\end{thebibliography}
